\documentclass[review]{elsarticle}
\usepackage{caption}
\usepackage{subcaption}
\usepackage{amsmath}
\journal{Nuclear Instrumentation Methods A}

\begin{document}

\begin{frontmatter}

\title{Design and studies of thick Gas Electron Multipliers fabricated in India}

\author[First]{Promita Roy\corref{mycorrespondingauthor}}
\address[First]{Saha Institute of Nuclear Physics, Kolkata-700064, a CI institute of Homi Bhabha National Institute, Mumbai-400094}

\cortext[mycorrespondingauthor]{Corresponding author}
\ead{promita.roy08@gmail.com}

\author[Second]{Purba Bhattacharya}
\address[Second]{School of Basic and Applied Sciences, Adamas University, Kolkata-700126}

\author[First]{Vishal Kumar}
\author[First]{Supratik Mukhopadhyay}
\author[First]{Nayana Majumdar}
\author[First]{Sandip Sarkar\corref{myfootnote}}

\cortext[myfootnote]{Retired Professor}




\begin{abstract}
THick Gas Electron Multipliers (THGEMs) are robust and high gain Micro Pattern Gaseous Detectors which are economically manufactured by standard drilling and etching of thin printed circuit boards. In this paper, we present our recent simulation as well as experimental studies on THGEMs which have been fabricated in India using local expertise. Two types of THGEMs have been fabricated; one set has holes without any external rim  and another set has holes with rims. These detectors have been characterized using argon-carbon dioxide and argon-isobutane gas mixtures. Electron transmission, effective gain, energy resolution and optimized working range studies have been presented for both the sets of THGEMs.

\end{abstract}


\end{frontmatter}


\section{Introduction}
The introduction of the Micro Pattern Gaseous Detectors(MPGDs) opened a new era of state-of-the-art detector technologies and they now represent the benchmarks for detector development in future accelerator-based experiments. Introduced by Fabio Sauli in 1997, the Gas Electron Multiplier (GEM) is a powerful addition to the family of fast radiation detectors~\cite{a0}\cite{a}. The THick Gas Electron Multiplier (THGEM) is a (geometrically) ten-fifteen fold expanded version of the standard Gas Electron Multipliers. It was first introduced at the Weizmann Institute of Science~\cite{a2}. This version of GEM is robust and comparatively easier to fabricate. Typically, THGEMs are made of FR4 PCB which have thickness in the range 0.2-1mm and have holes of few 100s of micron diameter. Multiplication of primary electrons occurs within these holes on irradiating the detector with a radiation source. These detectors offer high readout gain, typically in the range of 10$^3$-10$^5$ depending on the gas mixture~\cite{a3}\cite{a4}. In recent years, they have been widely used in large experiments such as dark matter measurement~\cite{b}, Cherenkov radiation imaging~\cite{c}\cite{c1}, and x-ray and neutron detection~\cite{d}. Apart from traditional FR4 THGEMs, a very recent paper~\cite{d1} reports the fabrication of glass THGEMs and their characterization in gaseous argon TPC. This paper explores circular holes as well as hexagonal holes. There has been quite a few simulation studies as well on the working of THGEMs~\cite{d2}\cite{d3}\cite{d4}.
Performance of THGEMs depend on various factors including its geometrical and electrical configurations.  The size of the etched rim around the THGEM holes, is essential for reducing discharge-occurrence probability significantly. This permits operation at higher THGEM voltages and hence yields higher detector gains~\cite{e}. Drift, transfer and induction field values are also found to affect gain and energy resolution, apart from applied voltage across THGEMs~\cite{f}. With hole size of few 100s of micron (100-200$\mu m$), they are found to have sufficiently good position resolution (close to 100$\mu m$) for muon tracking and imaging\cite{g}.

In the process of our search for a robust detector with decent spatial resolution, THGEMs with different geometrical parameters have been simulated and subsequently two types of THGEMs have been designed and fabricated and their experimental characterization have been carried out in different gas mixtures. The results have been compared with a CERN made THGEM. Section 2 provides the details of simulation of THGEMs with different geometrical parameters. This is followed by experimental studies with CERN-THGEM and locally fabricated THGEMs in section 3 and 4.

\section{Simulation study}

\begin{figure}[htbp]
	\centering
	\includegraphics[height=6.0cm]{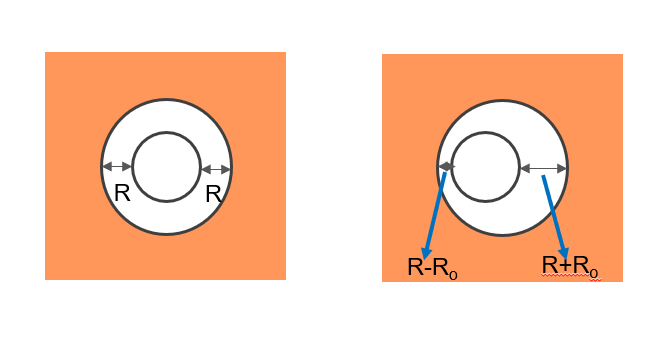}
	\caption{Left: THGEM with no misalignment of hole-rim( rim-offset is zero), Right: THGEM with non-zero misalignment; $R_o$ is the rim-offset.}
	\label{RIM offset}
\end{figure}

Prior to the design and fabrication of THGEMs locally, different hole geometries of THGEMs have been chosen keeping the dielectric thickness constant and numerical simulation has been done to have an estimate of the electron transparency and effective gain. Holes with different rim sizes have been simulated to see the effect of rim on electron collection and transmission. Moreover, to take into account the slightest of manufacturing defects such as misaligned rims, another set of simulation study has been performed varying the rim-offset as illustrated in figure\ref{RIM offset}.

\begin{table}[htbp]
	\caption{\textbf{Parameters used for GARFIELD simulation}}
	\label{THGEM_sim}
	\centering
	\small
	\begin{tabular}{ |c|c|c|c|c|c| } 
		\hline
		\bf{Study} & \bf{Thickness} & \bf{Hole radius} & \bf{Pitch} & \bf{Rim} & \bf{Rim-offset} \\ 
		& (d) & (r) & (p) & (R) & ($R_o$) \\
		--- & $\mu m$ & $\mu m$ & $\mu m$ & $\mu m$ & $\mu m$ \\
		\hline
		Rim size & 360 &  270 & 1020 & 0 - 125 & 0 \\
		fig:\ref{a1} & & & & & \\
		\hline
		Rim offset & 360 &  270 & 1020 & 120 & 0 - 120  \\ 
		fig:\ref{b1} & & & & & \\
		\hline
		Extreme cases & 360 & 270 & 1020 & 0, 120 &  0 \\
		fig:\ref{Rimsize-VGEM} & & & & & \\
	   \hline
	\end{tabular}
\end{table}

\begin{figure}[htbp]
	\centering
	\includegraphics[height=6.5cm]{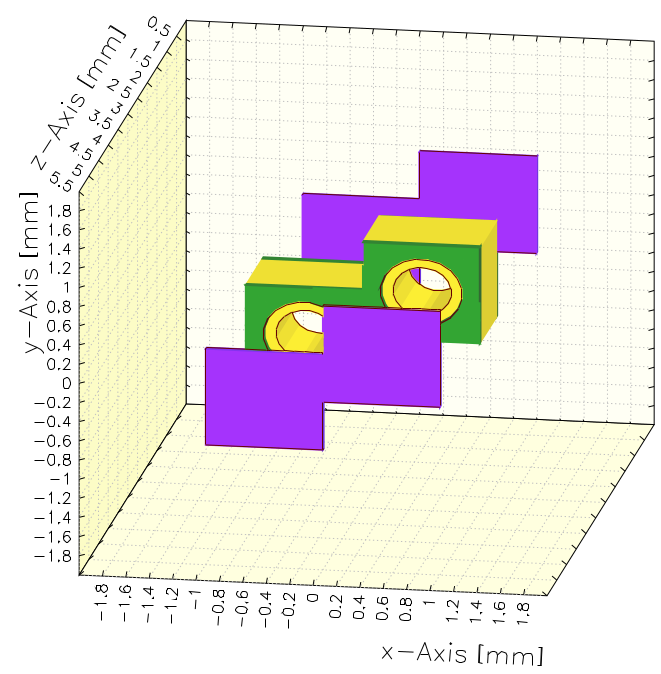}
	\caption{Straight THGEM holes arranged in staggered fashion}
	\label{THGEMgeo}
\end{figure}

The free, open source simulation framework of GARFIELD\cite{h} is used for the simulating the detector response. It simulates the primary ionization using HEED\cite{i} and electron-ion transport properties using MAGBOLTZ\cite{j}.
 The simulated THGEM has straight cylindrical holes arranged in hexagonal pattern as shown in figure~\ref{THGEMgeo}. The drift and induction gaps are chosen to be 1.5 cm and 3 mm respectively. Single electrons released from a height of 1.3 cm above the THGEM foils have been simulated. 10,000 such single electron events have been considered for each of the studies. Argon and isobutane gas mixture in the volumetric ratio of 95:5 has been used. Electric field maps have been computed using neBEM, which is again free and open source\cite{k}\cite{l}. Variation in electric field, collection, extraction and transmission efficiencies have been studied for different sets of THGEMs.

Table\ref{THGEM_sim} shows the various parameters used for the simulation study. THGEMs with no rim to THGEMs with rims as large as 125$\mu m$ have been simulated considering them to be perfectly aligned ($R_o$ = 0). Similarly, electron collection and transmission efficiencies have been simulated keeping the rim size of THGEMs fixed at 120$\mu m$ and varying the extent of misalignment from perfectly aligned holes to a maximum misalignment of 120$\mu m$.

\begin{figure}[htbp]
	\centering
	\begin{subfigure}{0.49\textwidth}
		\centering
		\includegraphics[width=\textwidth]{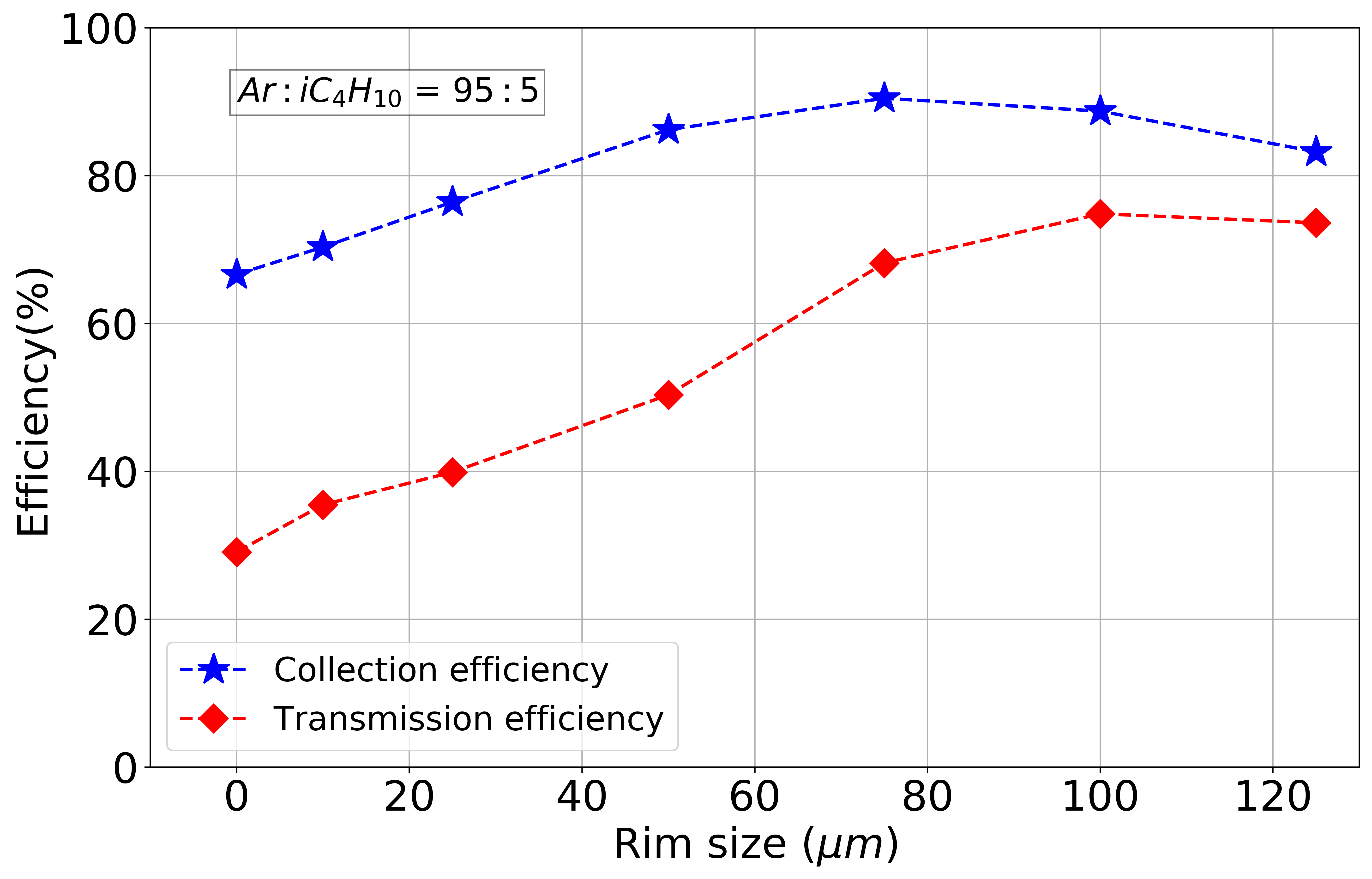}
		\caption{Different sizes of rim}
		\label{a1}
	\end{subfigure}
\hfill
	\begin{subfigure}{0.49\textwidth}
		\centering
		\includegraphics[width=\textwidth]{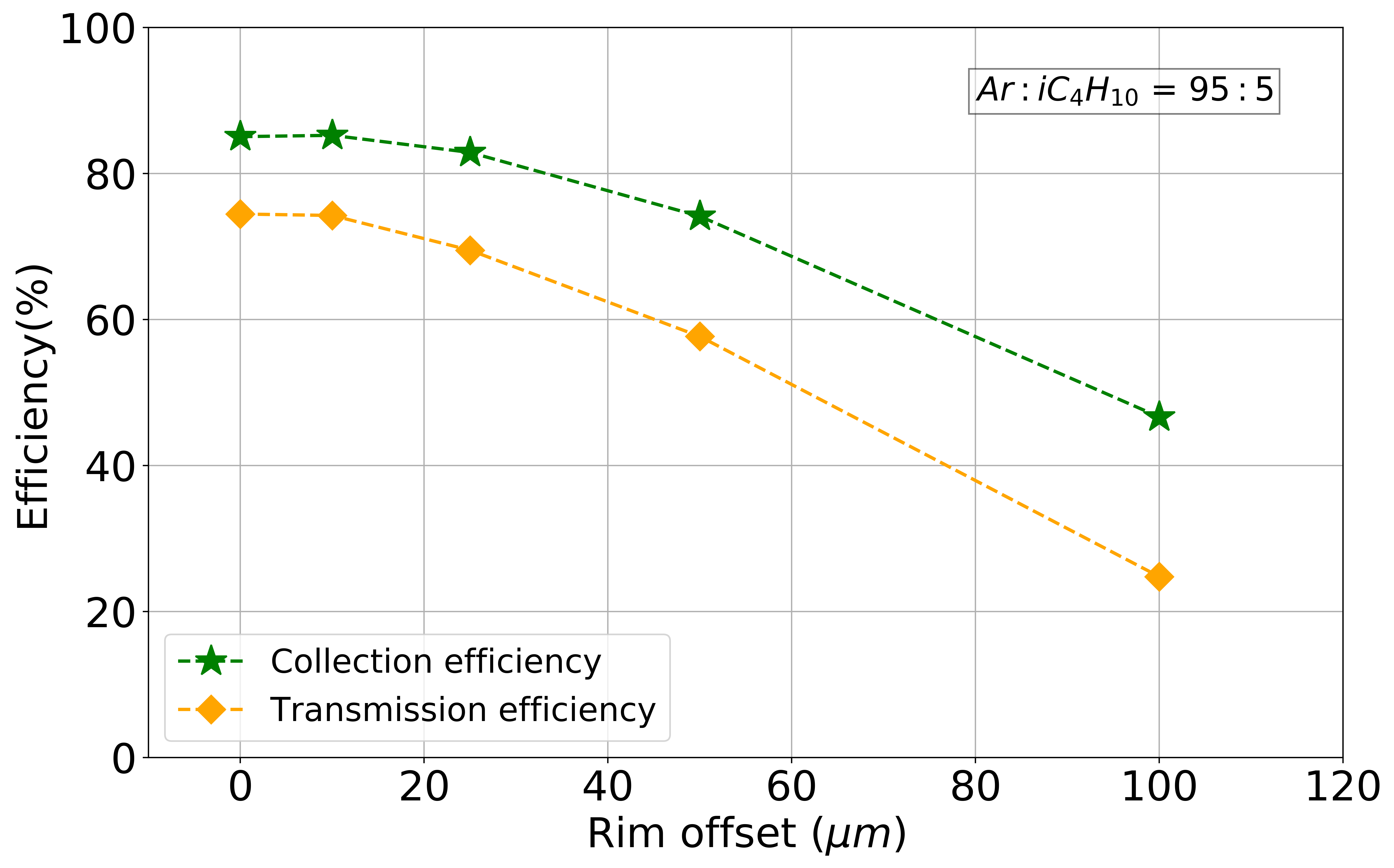}
		\caption{Different offsets in rim}
		\label{b1}
	\end{subfigure}
	\caption{Simulated results showing variation of collection and transmission efficiency with different (a) rim-sizes and (b) rim-offsets in Argon-isobutane mixture at $\Delta VGEM$ = 1050V}
	\label{Rimsize-offset}
\end{figure}

Transmission efficiency determines the effective gain and is defined as the number of electrons reaching the anode per unit primary electron in the drift gap~\cite{m}. Figure~\ref{Rimsize-offset} shows the simulated collection and transmission efficiency for THGEMs of different geomtries. Figure~\ref{a1} shows the variation of collection and transmission efficiency with increasing size of the rims. It is observed that both collection and transmission efficiency increases with increase in rim size. Beyond the rim size of 80$\mu m$, collection efficiency starts decreasing whereas transmission efficiency is found to almost saturate. Likewise, figure~\ref{b1} shows the variation in efficiencies with the increasing hole-rim misalignment. It is found that with increasing value of $R_o$ both the collection and transmission efficiency decreases. A close look at the values shows that an offset of 10 $\mu m$ can be allowed as there is hardly any change observed in the efficiencies.

\begin{figure}[htbp]
	\centering
	\begin{subfigure}{0.49\textwidth}
		\centering
		\includegraphics[width=\textwidth]{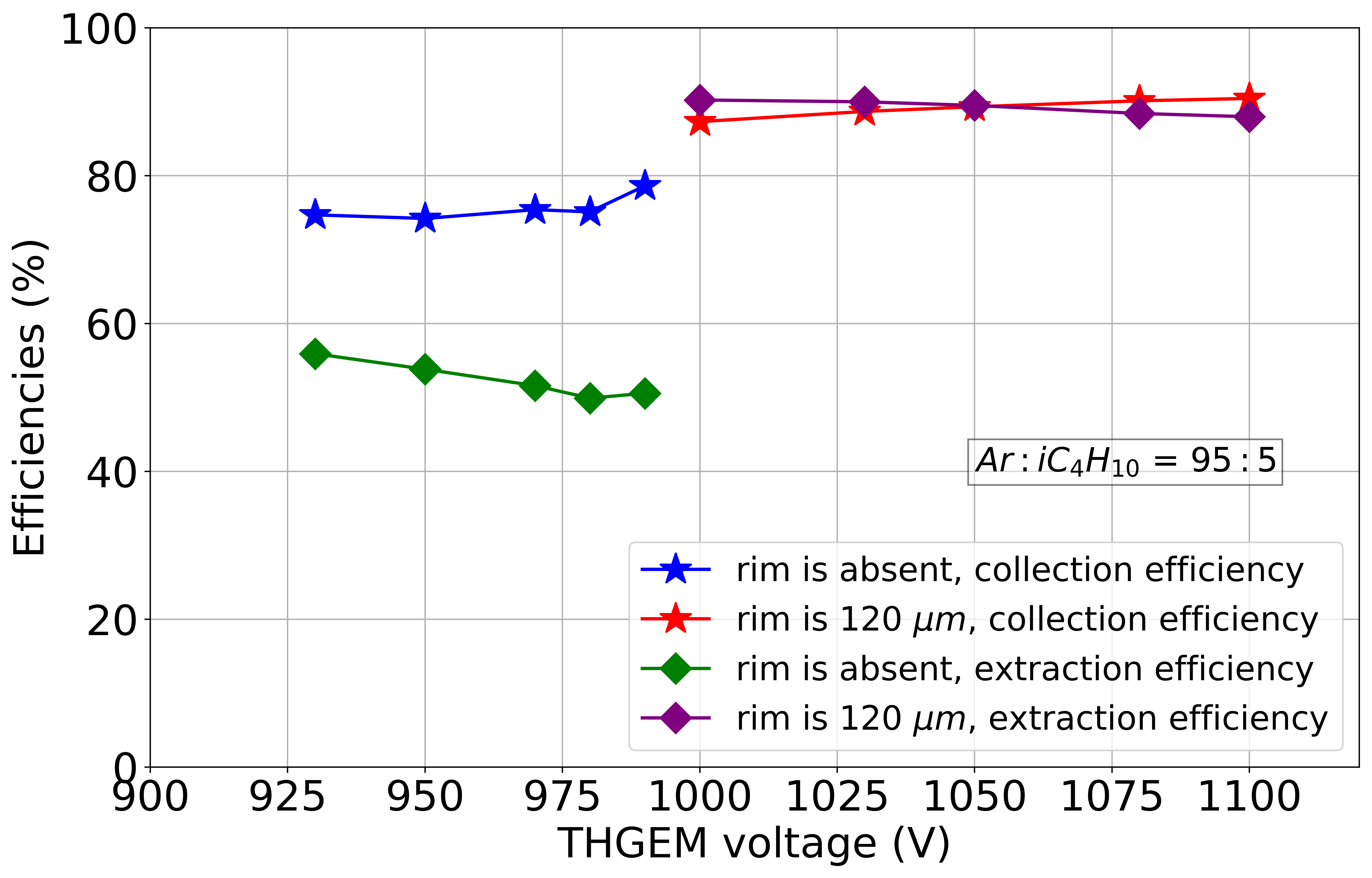}
		\caption{}
		\label{norim}
	\end{subfigure}
	\hfill
	\begin{subfigure}{0.49\textwidth}
		\centering
		\includegraphics[width=\textwidth]{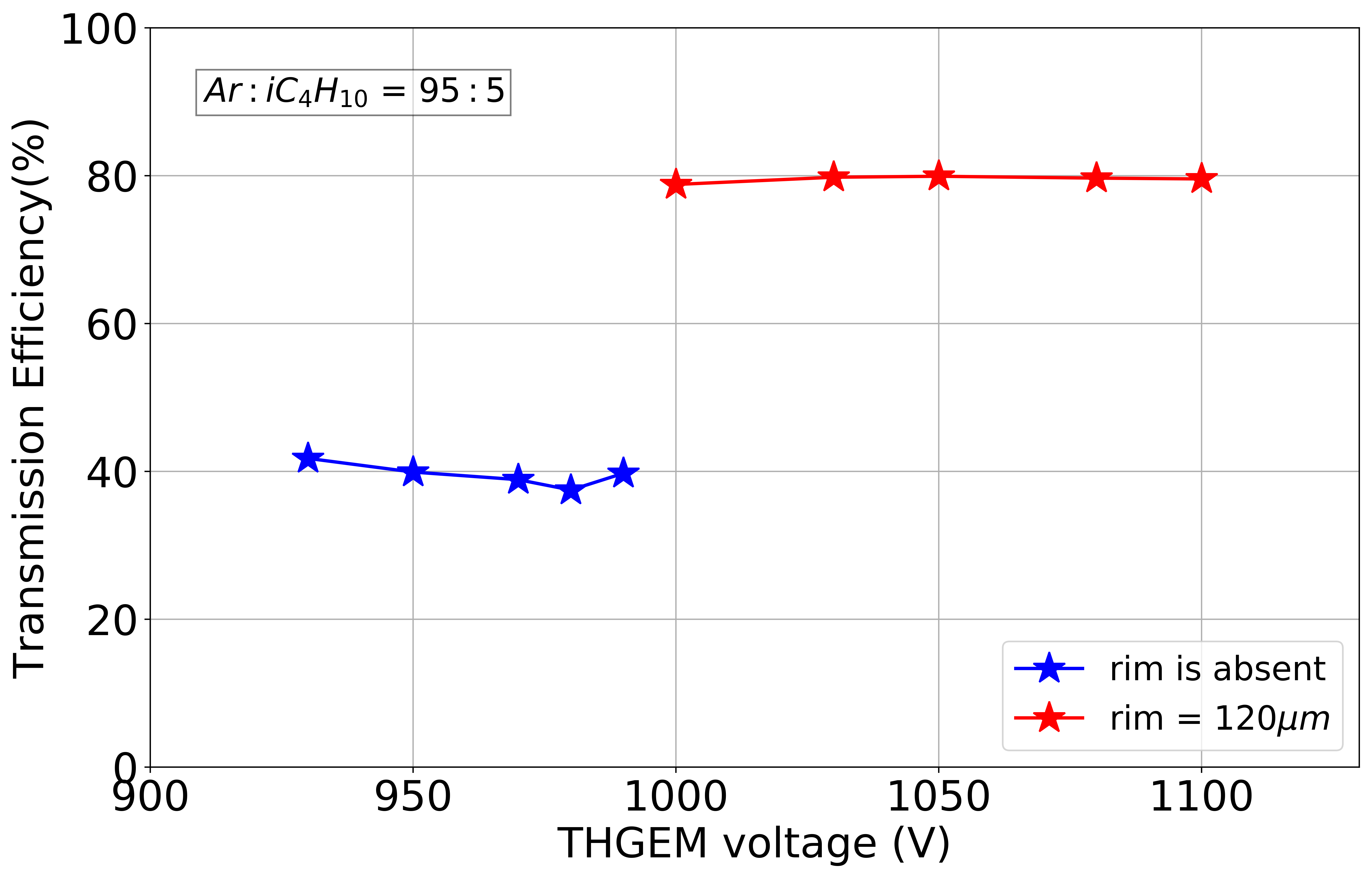}
		\caption{}
		\label{withrim}
	\end{subfigure}
	\caption{Simulated results showing variation of (a) collection, extraction efficiency and (b) transmission efficiency with THGEM voltages in Argon-isobutane mixture}
	\label{Rimsize-VGEM}
\end{figure}

As a next step of the study, THGEMs with two extreme rim sizes; one with no rim, another with rim size 120$\mu m$ have been simulated using 1cm $^{55}Fe$ tracks and variation of different efficiencies with applied THGEM voltage have been studied as shown in figure~\ref{Rimsize-VGEM}.

It has been observed that collection and extraction efficiencies for the one with rim of 120$\mu m$ is around 90\% throughout the working voltage range, whereas these efficiencies are quite less for the one with no rim(figure\ref{norim}). Absence of rim in the latter results in significant number of electron losses on top and bottom copper electrodes. This reduces the overall transmission efficiency as compared to the one with rim(figure\ref{withrim}).

\section{Experimental set-up and fabricated THGEMs}
\begin{figure}[htbp]
	\centering
	\includegraphics[width=6cm]{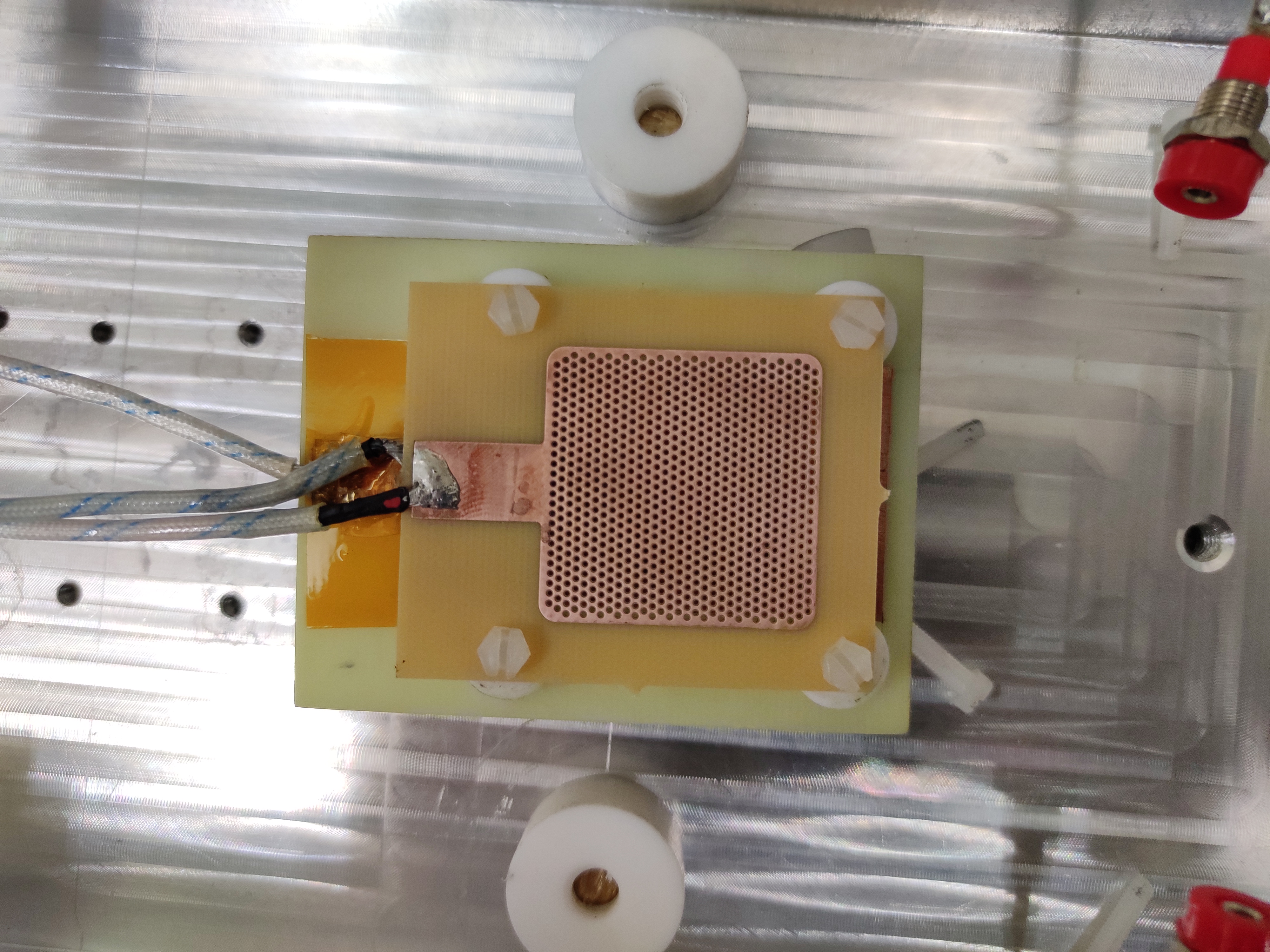}
	\caption{THGEM PCB from CERN ($TGC_{R}$) }
 \label{GEM_geometry_cern}
\end{figure}

A 40$\times$48 mm FR4 PCB based thick GEM manufactured in CERN ($TGC_{R}$) has been used for the initial testing and characterization. This PCB based detector is 800$\mu$m thick with around 50$\mu$m of copper coating on each side. It has straight cylindrical drills of diameter 500$\mu$m and rim of 100$\mu$m width arranged in hexagonal pattern. Figure~\ref{GEM_geometry_cern} shows the picture of the $TGC_{R}$ .

\begin{figure}[htbp]
\centering
  \includegraphics[height=5cm]{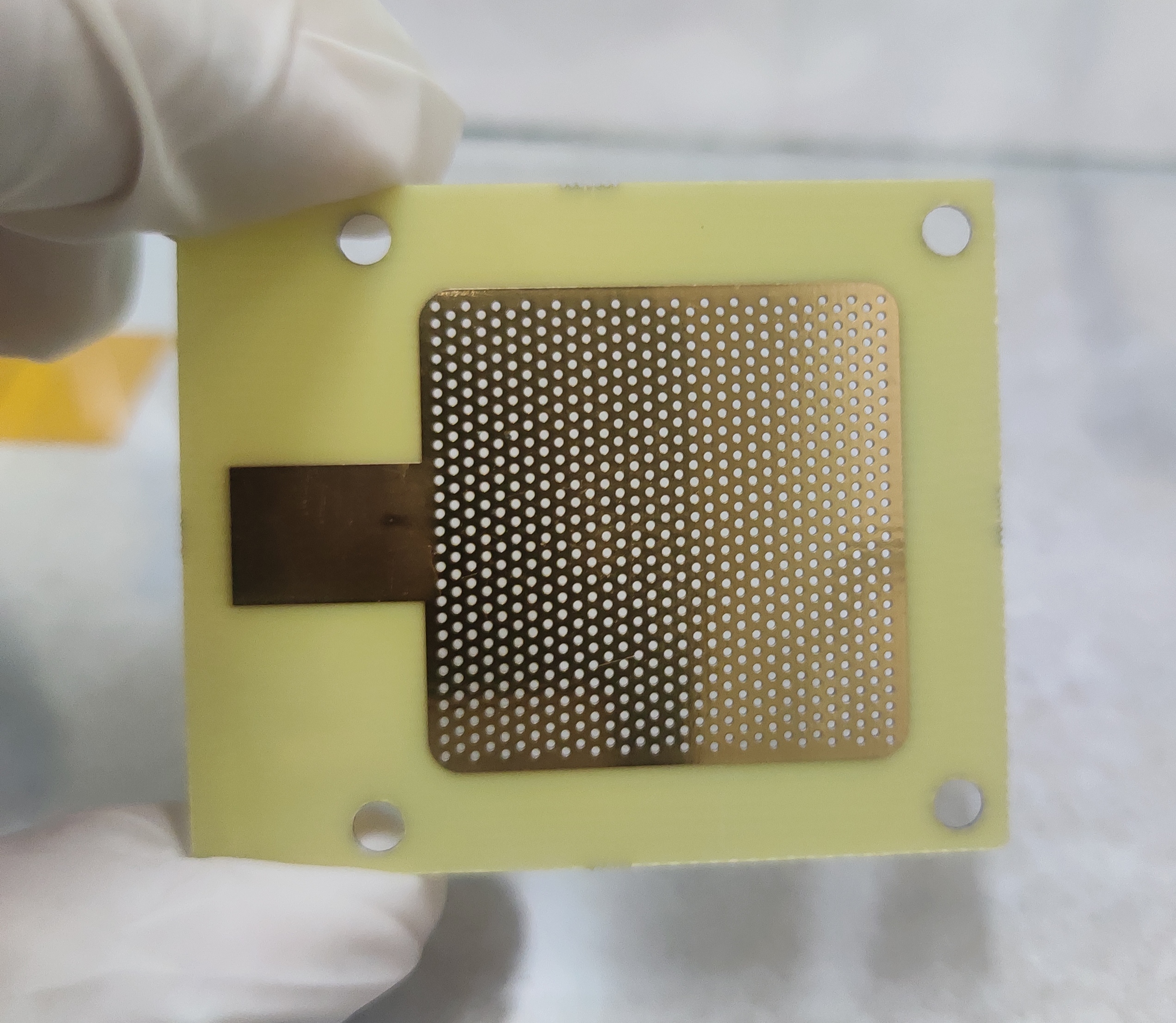}
  \includegraphics[height=5.0cm]{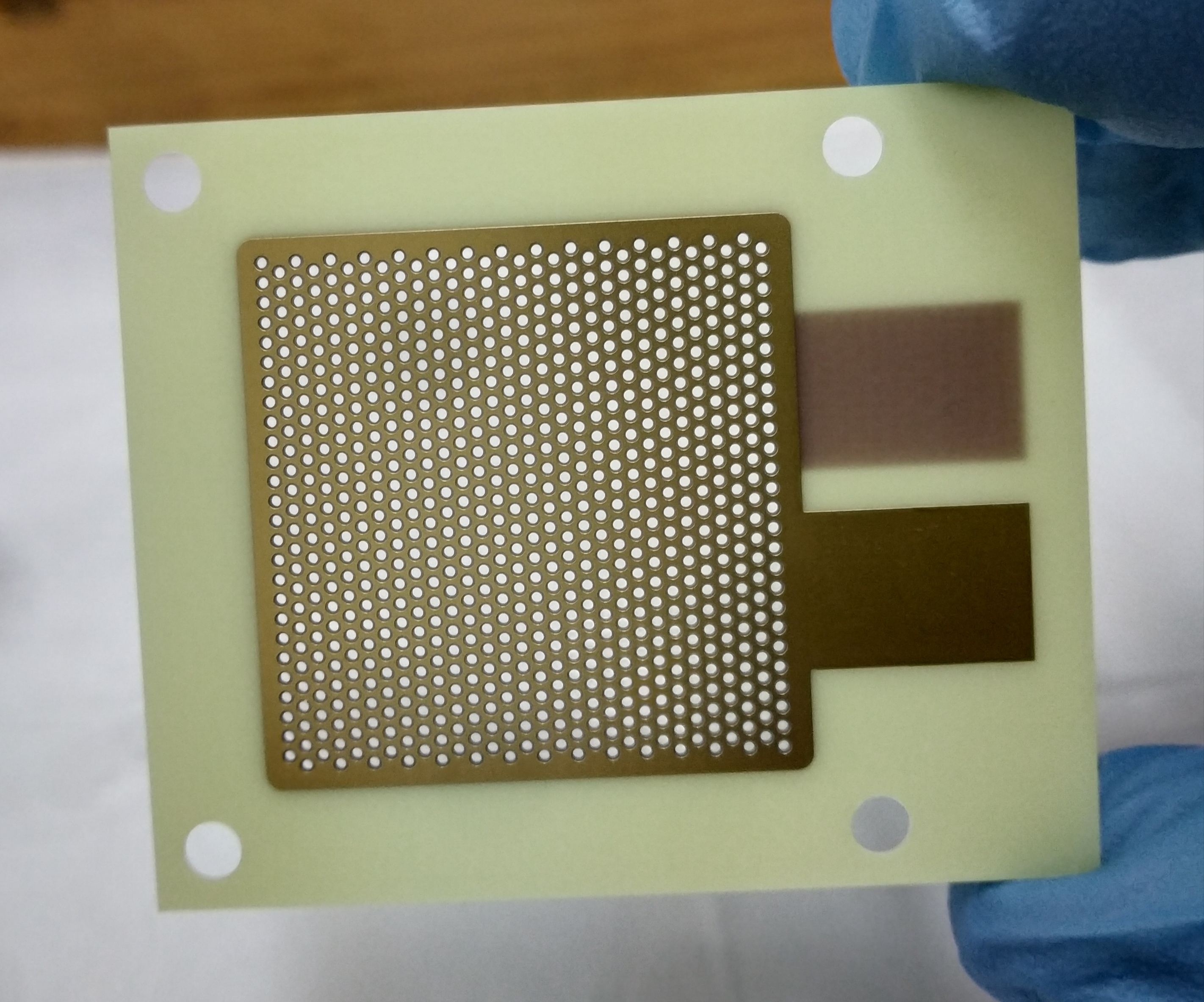}
\caption{Locally fabricated THGEMs $TGS$ (left) and $TGS_{R}$ (right)}
\label{fig:wo_rim}
\end{figure}

Following the conclusion drawn from the simulation results in section 2, another two sets of THGEMs of same size as $TGC_{R}$ (40$\times$48 mm) with($TGS_{R}$) and without rim($TGS$)have been designed at SINP (Saha Institute of Nuclear Physics) using AutoCAD software. Table~\ref{THGEM_config} gives the details of the THGEM geometries used for experiment. Abbreviations $TGC_{R}$, $TGS$ and $TGS_{R}$ have been used for these three THGEMs where C and S stands for CERN and SINP respectively. 'R' denotes the presence of rim. The local THGEMs have been fabricated by drilling straight holes in FR4 based printed circuit boards. The chosen FR4 is based on IS410 Lead-free Epoxy Laminate and Prepreg PCB material. These FR4-PCBs are 370-380$\mu$m thick with nearly 50$\mu$m copper coating on each side. Holes are arranged in hexagonal pattern similar to $TGC_{R}$ to increase the optical transparency of the foils. Both the sets of copper coated FR4-PCBs have holes of diameter 500$\mu$m and pitch of 1mm. However, just one set($TGS_{R}$) has 100$\mu$m wide rims around the holes, while the other set($TGS$) has no rim around the holes. The average hole-rim misalignment in these locally fabricated THGEMs is around 20$\mu m$. Figure~\ref{fig:wo_rim} shows the locally fabricated THGEMs ($TGS$)  and ($TGS_{R}$) respectively.

\begin{table}[htbp]
	\caption{\textbf{Different geometries of THGEMs used}}
	\label{THGEM_config}
	\centering
	\small
	\begin{tabular}{ |c|c|c|c|c|c| } 
		\hline
		\bf{THGEM} & \bf{Type} & \bf{Thickness} & \bf{Hole radius} & \bf{Pitch} & \bf{Rim}\\ 
		 \bf{no.}& & (d) & (r) & (p) & (R)\\
		--- &  --- & $\mu m$ & $\mu m$ & $\mu m$ & $\mu m$ \\
		\hline
		$TGC_{R}$ & CERN THGEM with rim  & 800 &  250 & 1000 & 100 \\
		\hline 
		$TGS$ & SINP THGEM without rim & 370-380 &  250 & 1000 & none \\ 
		\hline
		$TGS_{R}$ & SINP THGEM with rim & 370-380 &  250 & 1000 & 100 \\ 
		
		\hline
	\end{tabular}
\end{table}

\begin{figure}[htbp]
	\centering
	\includegraphics[width=8.05cm]{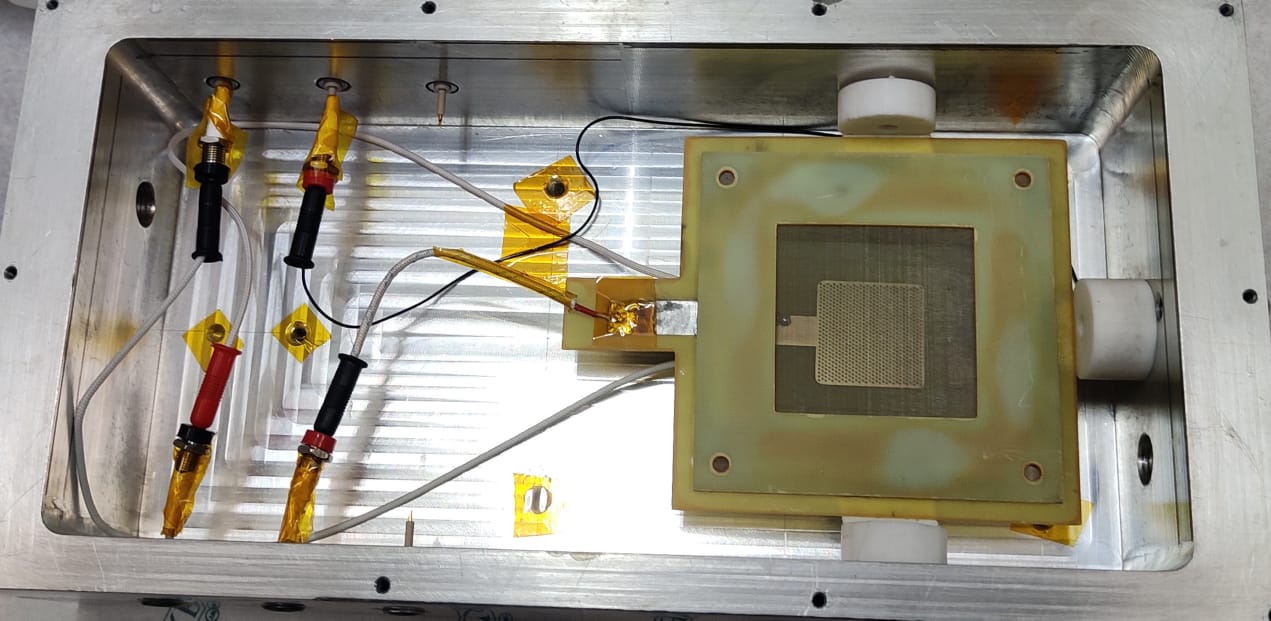}
	\caption{Aluminium test box containing the THGEM}
	\label{Al_box}
\end{figure}

 A stretched mesh secured in a rectangular frame has been used as the drift cathode and and a copper plated PCB has been used as the readout anode. These electrodes along with the THGEM PCB have been kept inside an aluminium test box which has all the electrical connections as well as gas inlet and outlets as shown in the figure~\ref{Al_box}. Teflon holders and nylon screws have been used to secure the drift plate, THGEM and readout anode plate inside the test box. The chosen drift gap is 10mm and induction gap is 3mm. The lid of the test box has a mylar window to place the source and irradiate the detector.

\subsection{Measurement details and procedure}
\begin{figure}[htbp]
    \centering
    \includegraphics[width=11.0cm]{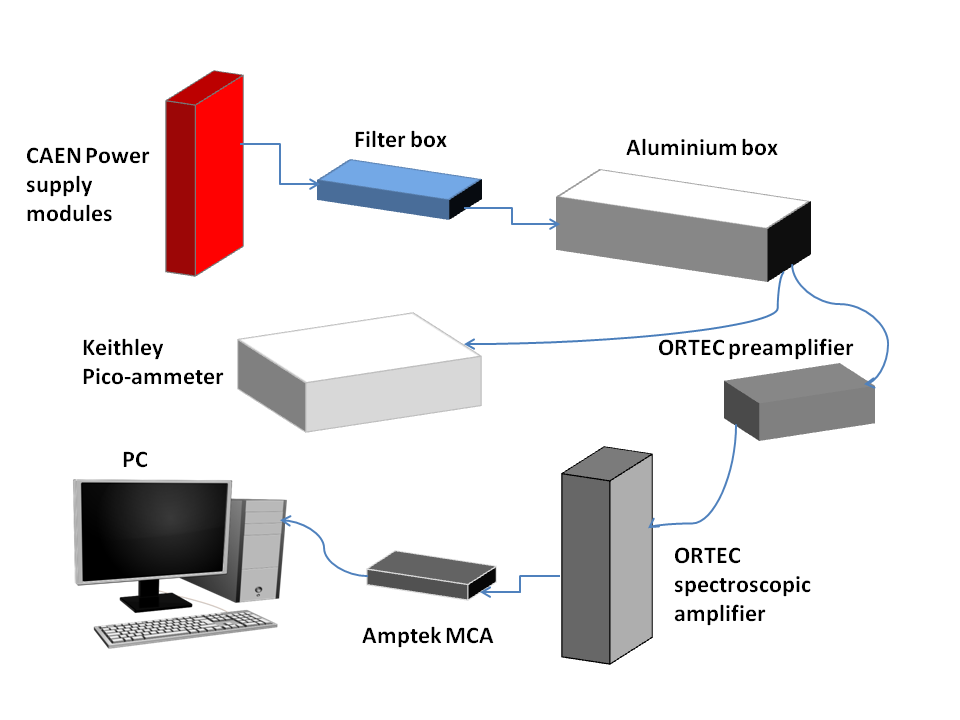}
    \caption{Experimental setup with THGEM}
    \label{exp_setup}
\end{figure}

Voltages to drift electrode and THGEM PCB are applied using CAEN power supply modules N1471 and N471A. Amptek pocket MCA has been used for getting the $^{55}Fe$ spectra and the readout current measurement has been carried out using Keithley 6487 pico-ammeter. Temperature, pressure and humidity have been constantly monitored using an Arduino based sensor. Ortec 142 IH preamplifier and Ortec 672 spectroscopy amplifier have been used for shaping and amplifying the readout signal. Figure~\ref{exp_setup} shows the typical experimental set up containing the test box, filter box, pre-amplifier, spectroscopy amplifier, MCA and PC. Data from pre-amplifier is fed to spectroscopic amplifier, which is then fed to the MCA interfaced to a desktop. Data from readout anode is also fed to the pico-ammeter as shown in the figure~\ref{exp_setup}.

\subsection{Pre-characterization procedures}
\begin{figure}[htbp]
	\centering
	\includegraphics[width=5.5cm]{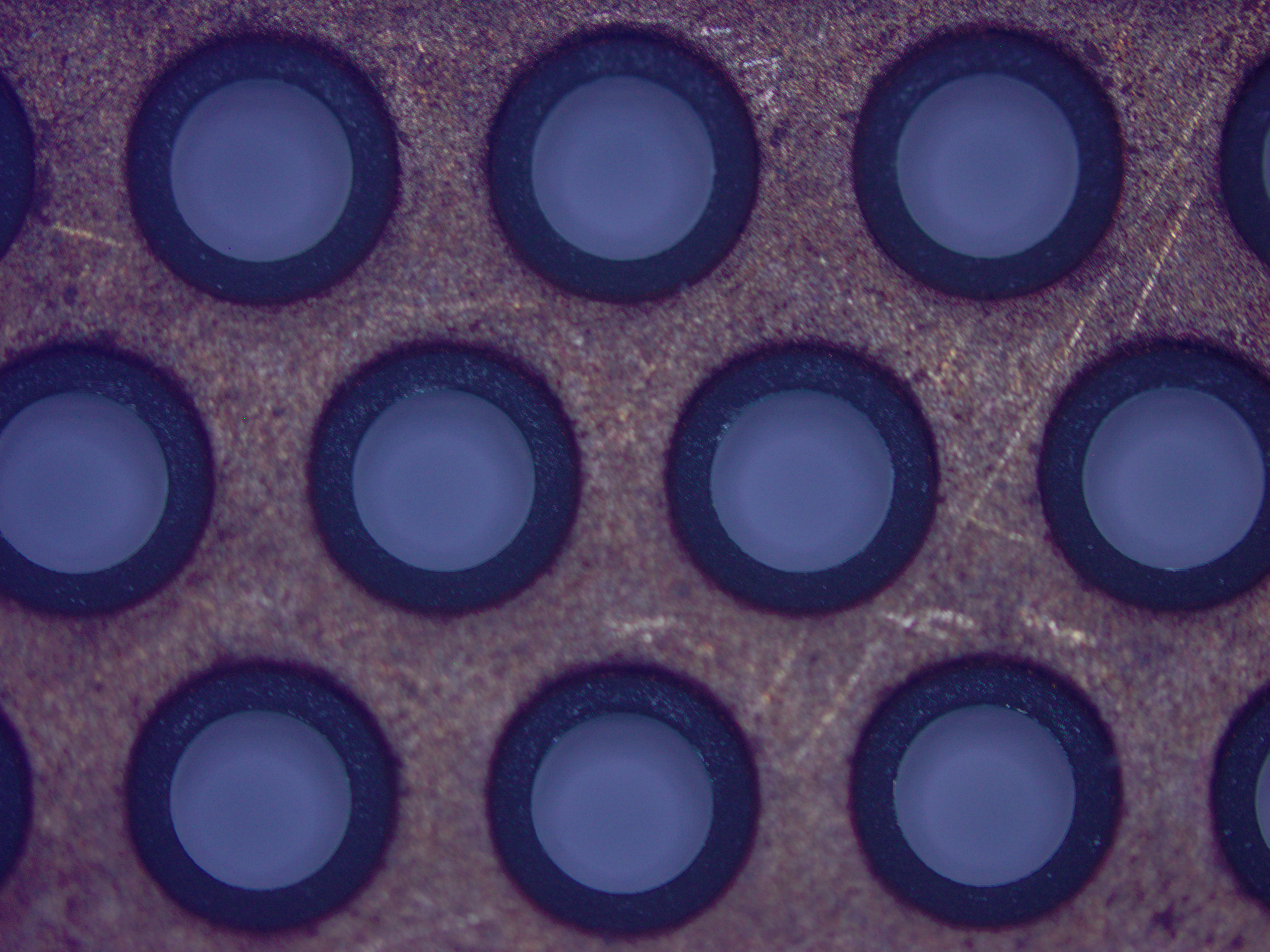}
	\centering
	\includegraphics[width=5.5cm]{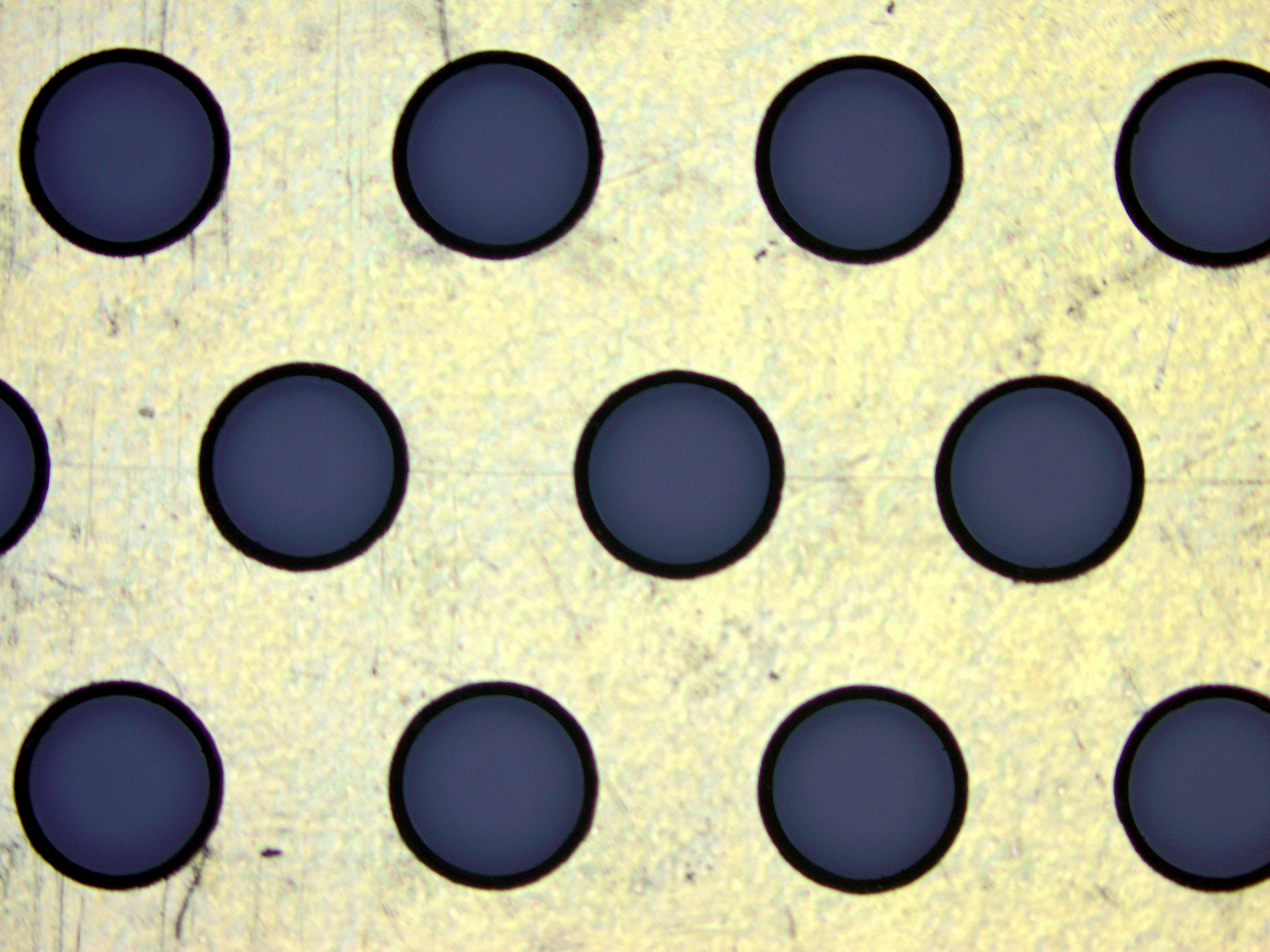}
	\includegraphics[width=5.5cm]{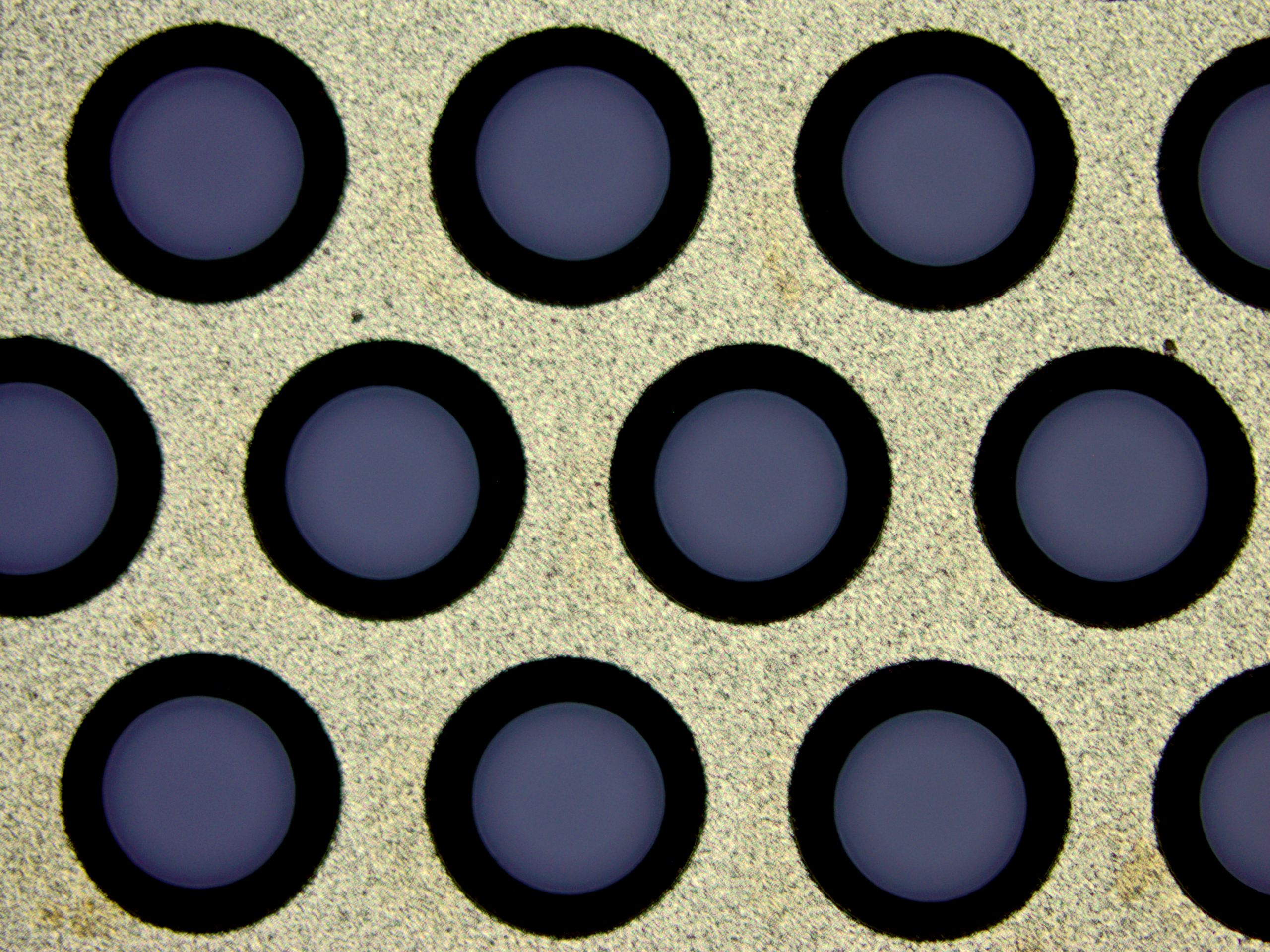}
	\caption{Microscopic images of $TGC_{R}$ (top left), $TGS$ (top right) and $TGS_{R}$ (bottom)}
	\label{fig:micro_image}
\end{figure}

Prior to the characterization of the THGEMs, resistivity of all the detectors (PCBs) have been checked and they have been found to have resistivity more than 11G$\Omega$ on application of 1000V. Microscopic images have been saved for all the three THGEMs as shown in figure~\ref{fig:micro_image}. Copper coating of $TGS$ and $TGS_{R}$ have 2-3$\mu$m thick gold plating to protect it from corrosion. Microscopic images have been used to cross-check the geometrical parameters of these locally fabricated THGEMs and they have been mentioned in table~\ref{THGEM_config}.

The next procedure was to clean the THGEM PCBs and that has been done using the conventional cleaning measures which include soaking the detectors in iso-propanol for hours and baking them in nitrogen gas with high flow-rate until the test box is flushed entirely around 5 times. Initial high voltage testing has been performed for all the three types of THGEMs using nitrogen gas, keeping in mind the breakdown limits from Paschen curve. These breakdown voltage limits have been found to be close to 2200V and 4000V for PCBs of thickness 370$\mu m$ and 800$\mu m$ respectively at atmospheric pressure. Argon-carbon dioxide gas mixture in the volumetric ratio 90:10 and argon-isobutane mixture in the ratio of 95:5 have been used for various measurements. These measurements have been carried out using a $^{55}$Fe extended source and a point source whose rate of radiation have been controlled by collimating the source using collimators having different apertures.

\section{Experimental results} 

All the three THGEMs ($TGC_{R}$, $TGS$ and $TGS_{R}$) have been characterized using two 5.9keV $^{55}$Fe sources of different rates in argon based gas mixtures. Extensive gain measurements have been performed for different applied voltages within the discharge limit. Table~\ref{Exp_config} lists the various experimental measurements and some of the important parameters used.
\begin{table}[htbp]
	\centering
	\caption{\textbf{Experimental studies}}
	\label{Exp_config}
	\begin{center}
		\small
		\begin{tabular}{ |c|c|c|c|c| } 
			\hline
			\bf{Section} & \bf{THGEM} & \bf{Measurement} & \bf{Gas mixture} & \bf{Source(kHz)} \\ 
			\hline
			4.1.1 & $TGC_{R}$ & Characterization & $Ar:CO_2$=90:10, & 1.2, 4.8 \\ 
			& & & $Ar:iC_4H_{10}$=95:5 & \\
			\hline
			4.1.2 & $TGS$, &  Characterization & -do- &1.8, 4.8\\ 
			&  $TGS_{R}$ & & & \\
			\hline
			4.2 & $TGC_{R}$, $TGS$, & Gain measurement & -do- &1.65 \\ 
			& $TGS_{R}$ & & & \\
			\hline
			4.3 & $TGC_{R}$, $TGS$,  & Energy resolution & -do- & 1.7\\
			& $TGS_{R}$ & & & \\
			\hline
		\end{tabular}
	\end{center}
\end{table}

\subsection{Characterization}
Characterization of THGEM includes getting $^{55}$Fe spectra for different rates of the source, studying the optimum working voltage range and optimizing the electron transparency to maximize the gain.
\subsubsection{$TGC_{R}$}

\begin{figure}[htbp]
    \centering
    \includegraphics[height=5.0cm]{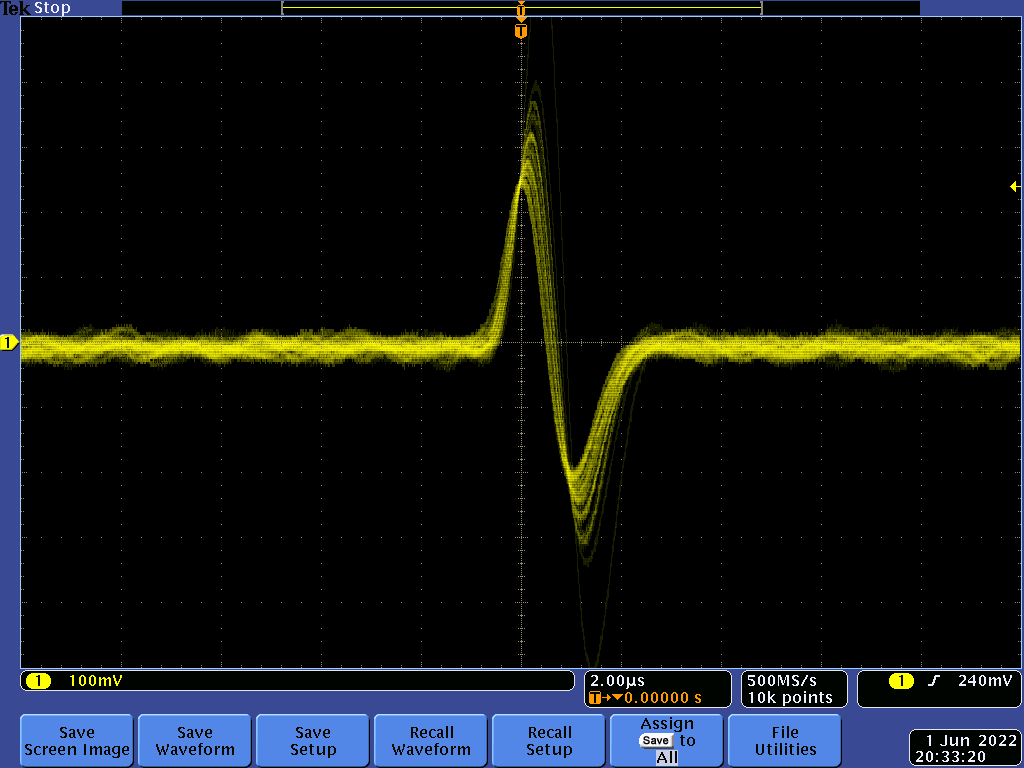}
    \includegraphics[height=6.0cm]{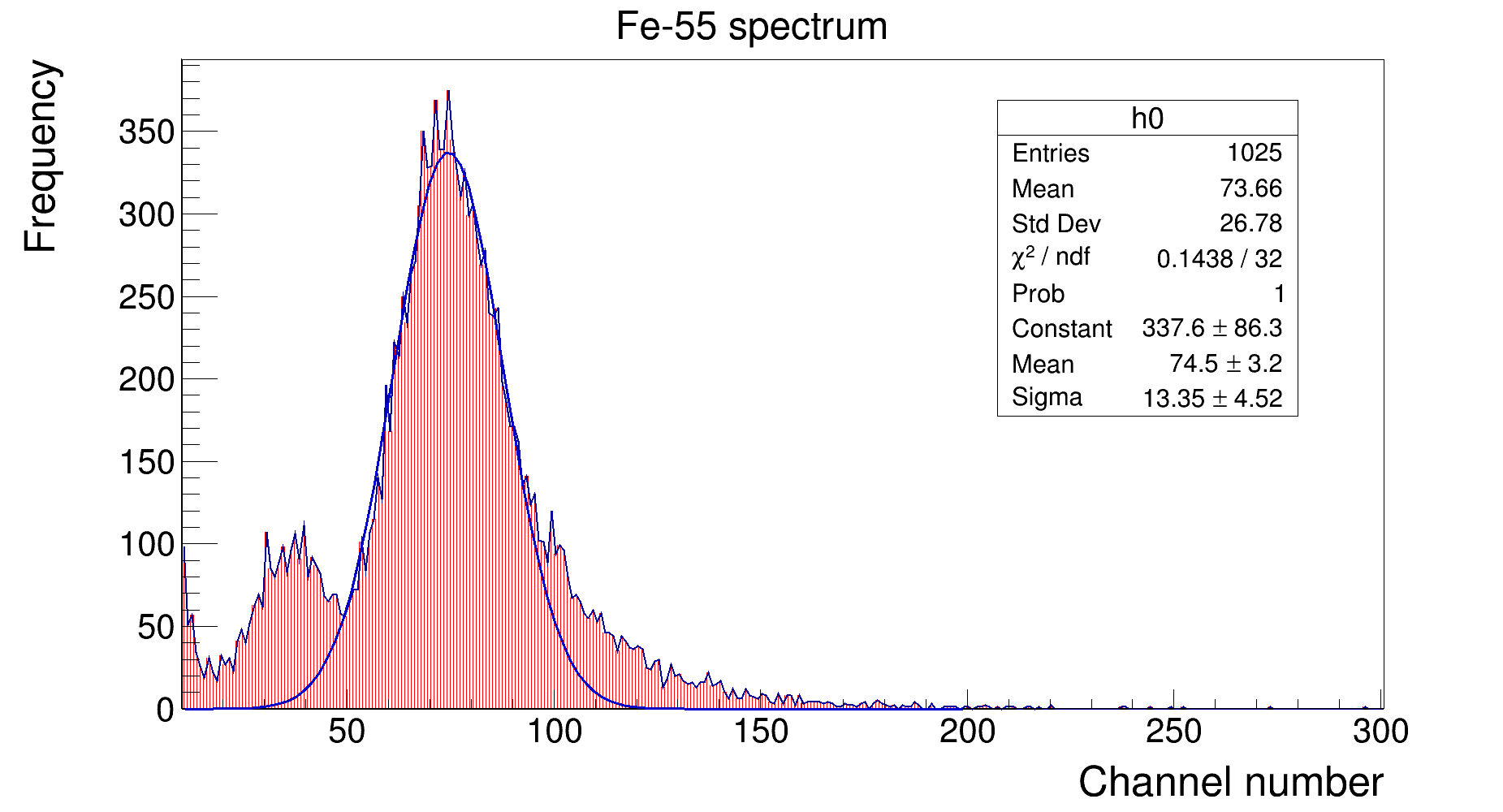}
    \caption{Oscilloscope signal(top) and $^{55}$Fe spectrum(bottom) obtained using $TGC_{R}$}
    \label{Fe-55cern}
\end{figure}

Figure~\ref{Fe-55cern} shows the output from the spectroscopic amplifier in the oscilloscope on irradiating the detector with a source as well as the fitted photopeak of the $^{55}$Fe spectrum from the MCA. Both the photopeak and Ar-escape peak look well separated.

    

\begin{figure}[htbp]
    \centering
    \includegraphics[width=10.05cm]{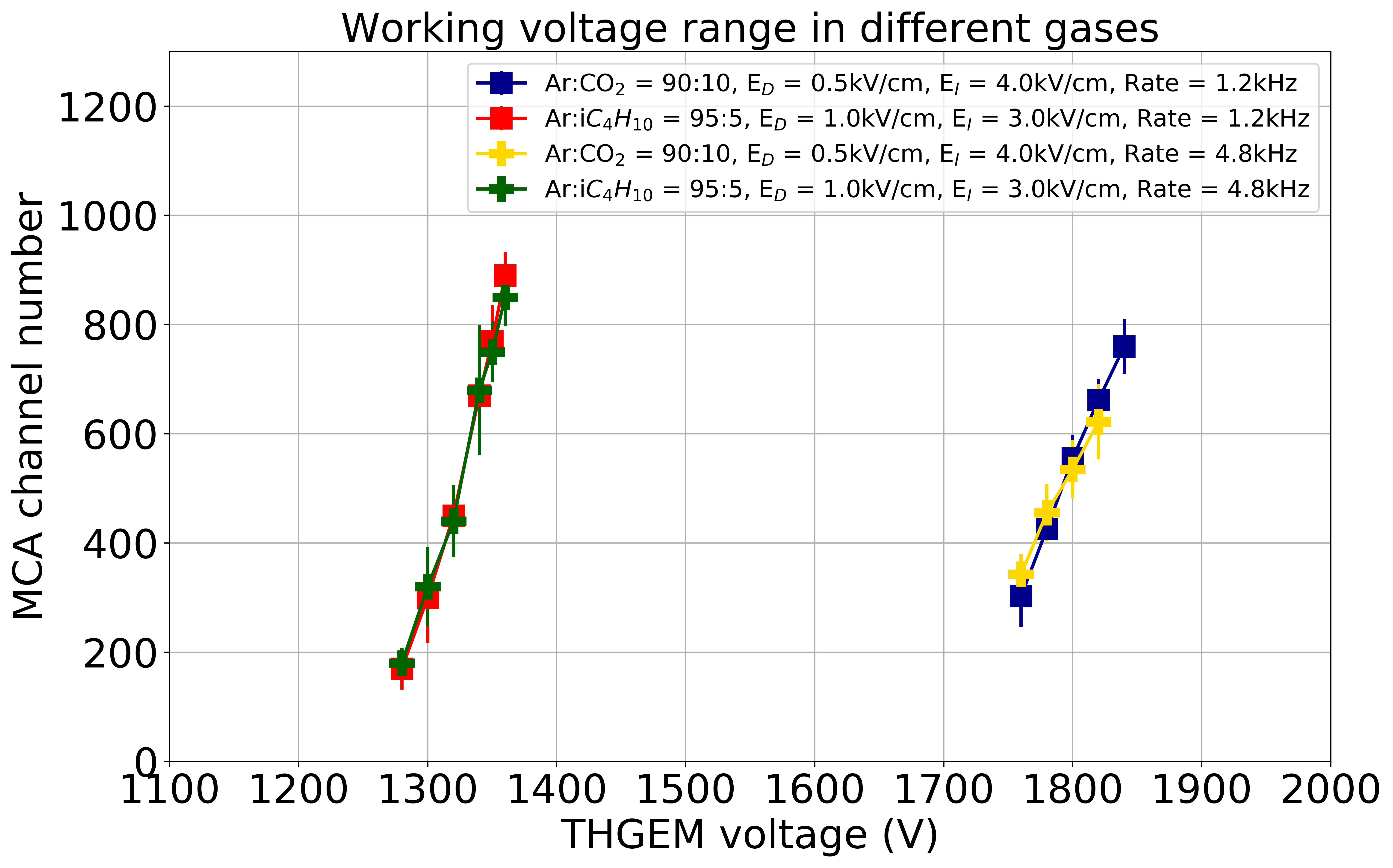}
    \caption{Operating range in argon based gas mixtures for $TGC_{R}$}
    \label{Oprange_cern}
\end{figure}

To study the optimum working range, two rates of $^{55}$Fe source have been used in two different argon based gas mixtures and corresponding MCA spectra have been recorded while varying the THGEM voltage. Drift and induction fields have been kept constant throughout the measurement. The photopeak of the obtained MCA spectra have been fitted with a gaussian function to get the mean photopeak channel number corresponding to each applied voltage. Figure~\ref{Oprange_cern} shows the operating voltage range for $TGC_{R}$ for two different rates (1.2kHz and 4.8kHz) in two different gas mixtures; Ar:$CO_2$ = 90:10 and Ar:$iC_4H_{10}$ = 95:5. The maximum attainable gain or MCA channel number was defined as the one at which micro-discharges were observed less than 1 per minute. Here, a micro-discharge has been defined as an event which yields current($I_{MON}$) more than 5nA. It has been noted that for a low-rate source, operating voltage range is larger for both the gases. Furthermore, the voltage range is larger in argon-isobutane mixture than in argon-carbon dioxide mixture due to the former having good discharge quenching property.

\begin{figure}[htbp]
	\centering
	\begin{subfigure}{0.49\textwidth}
		\centering
		\includegraphics[width=\textwidth]{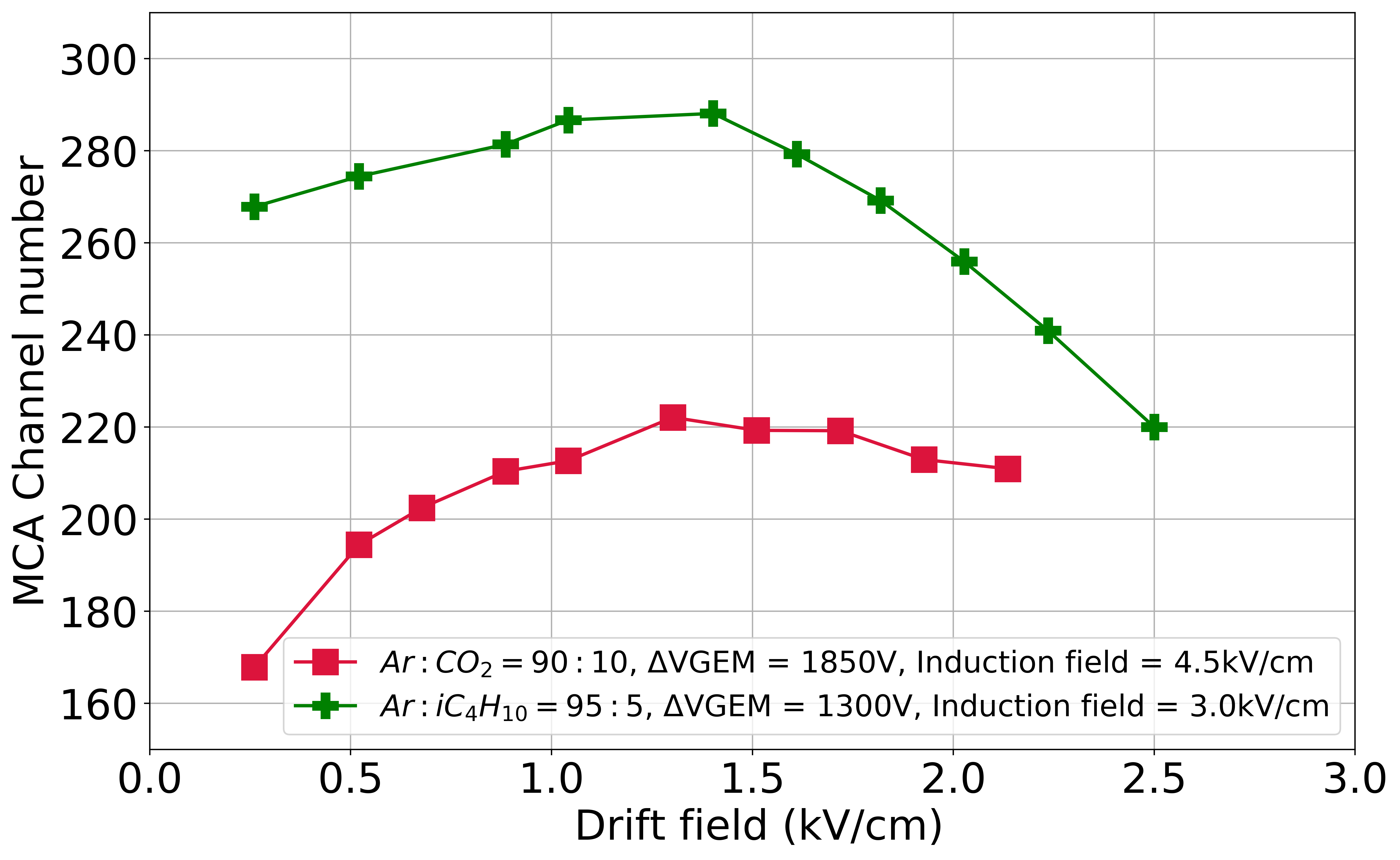}
		\caption{}
		\label{DF}
	\end{subfigure}
	\hfill
	\begin{subfigure}{0.49\textwidth}
		\centering
		\includegraphics[width=\textwidth]{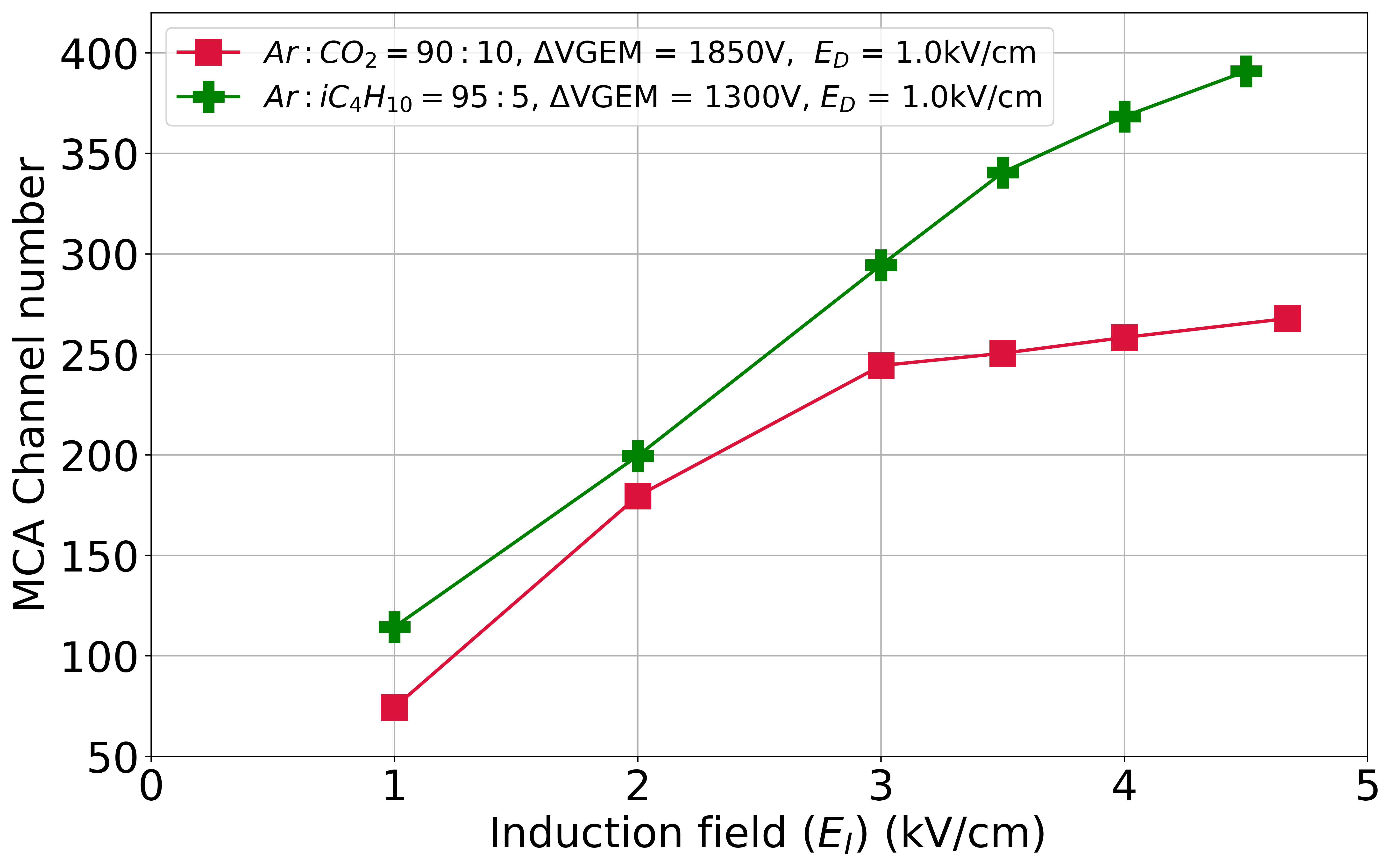}
		\caption{}
		\label{IF}
	\end{subfigure}
	\caption{Variation of gain for $TGC_{R}$ with (a) drift field and (b)induction field.}
	\label{drift_ind_cern}
\end{figure}

To optimize the drift ($E_D$) and induction field ($E_I$) values or to maximize the electron collection within the hole and electron transmission to the anode at a given voltage, MCA spectra have been saved by varying the drift and induction fields keeping the THGEM voltage constant. Figure~\ref{DF} shows the variation of mean MCA photopeak channel number with different values of $E_D$ keeping THGEM voltage and $E_I$ constant. It is observed that the mean channel number increases with the drift field, forms a plataeu-like region and finally decreases for both the gases. 

Similarly, figure~\ref{IF} shows the variation of mean photopeak MCA channel number for both the gases with different values of $E_I$ keeping $\Delta$VGEM and $E_D$ constant. It is observed that MCA channel number increases with increasing $E_I$.

\subsubsection{$TGS$ and $TGS_{R}$ }

The same procedure has been adopted to characterize the locally fabricated SINP THGEMs ($TGS$ and $TGS_{R}$). Details of the experimental cofiguration are given in the table~\ref{Exp_config}.

\begin{figure}[htbp]
	\centering
	\begin{subfigure}{0.49\textwidth}
		\centering
		\includegraphics[width=\textwidth]{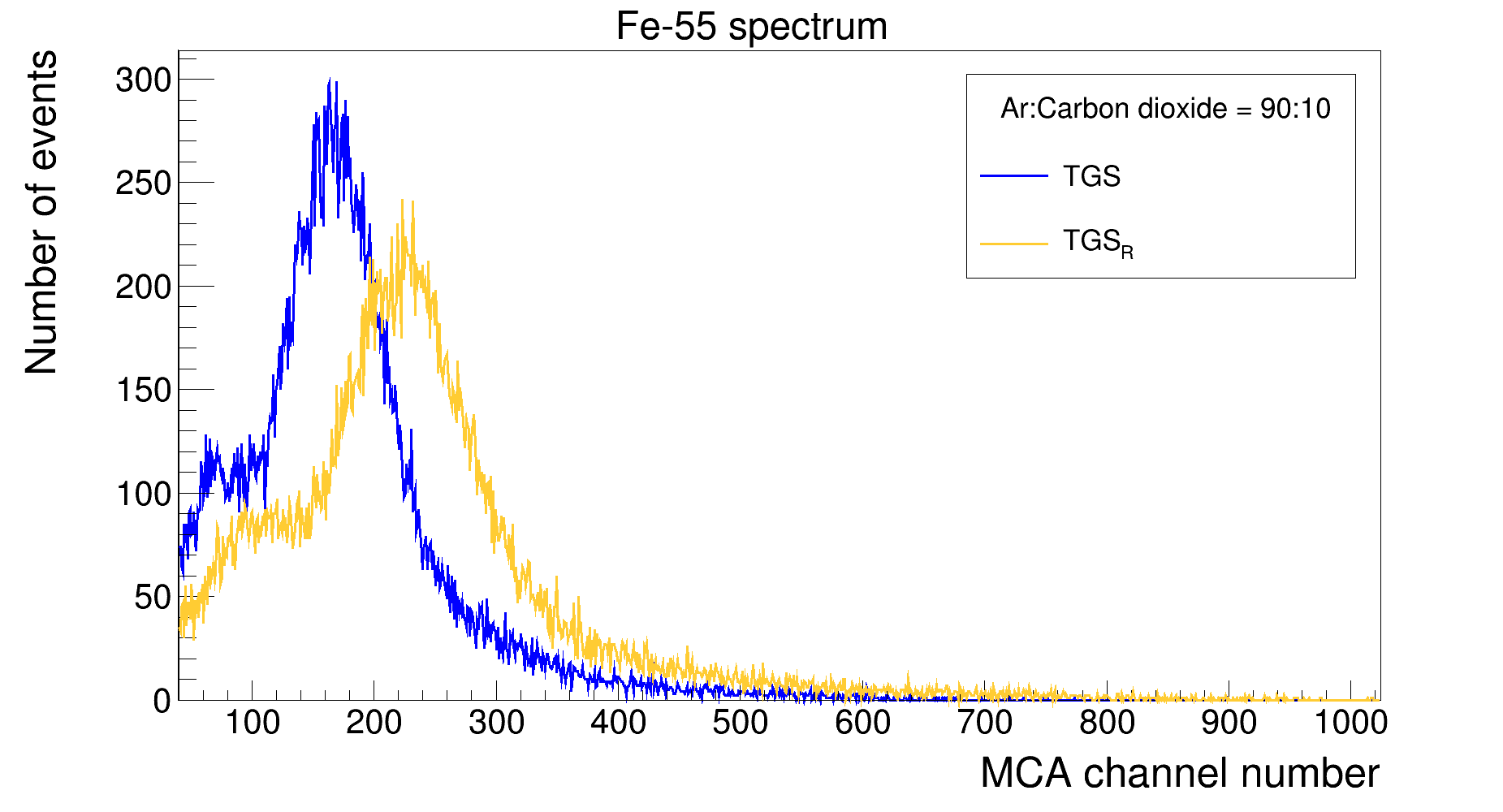}
		\caption{}
		\label{spec-9010}
	\end{subfigure}
	\hfill
	\begin{subfigure}{0.49\textwidth}
		\centering
		\includegraphics[width=\textwidth]{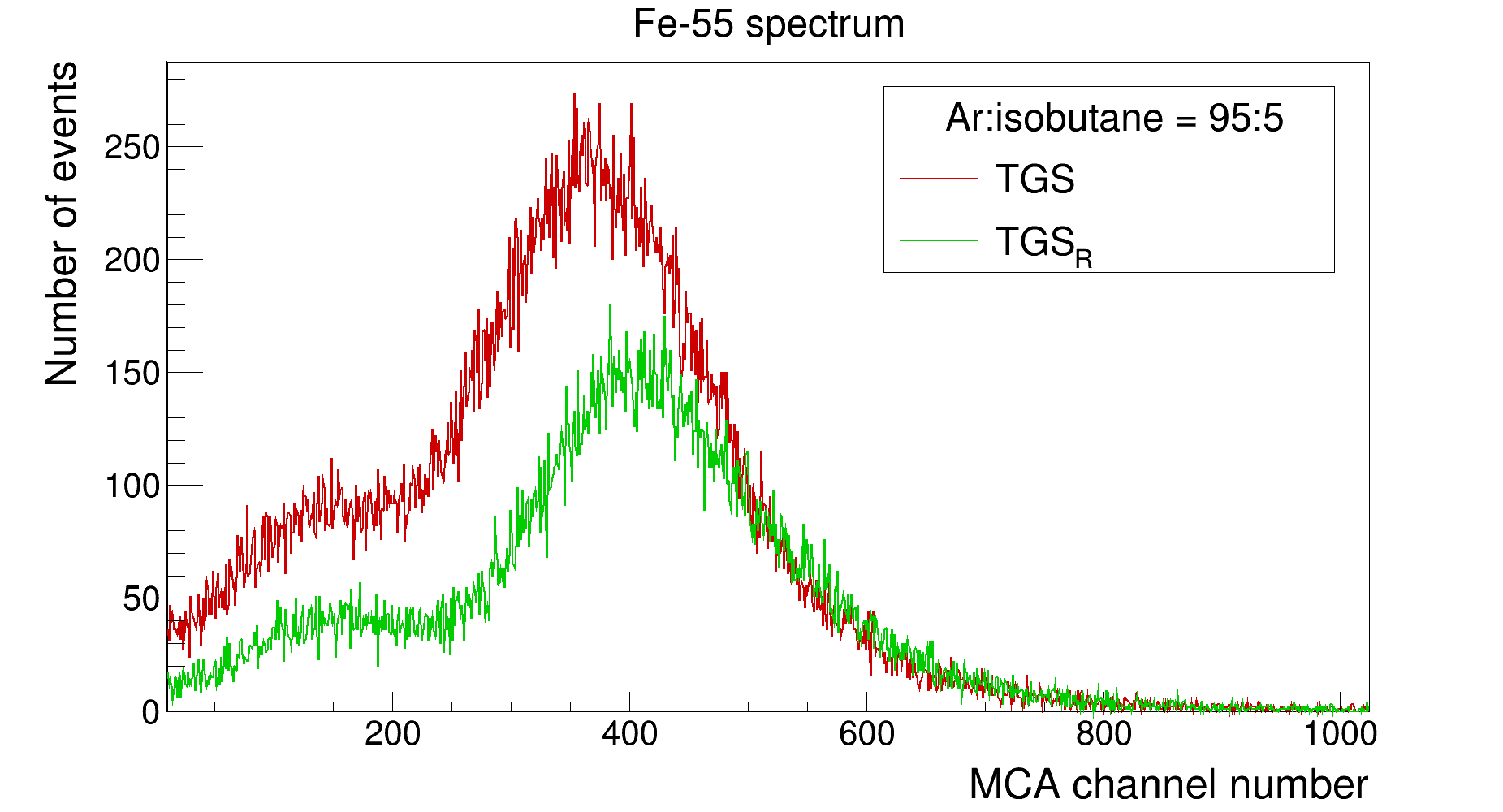}
		\caption{}
		\label{spec-955}
	\end{subfigure}
	\caption{Fe-55 spectrum obtained for $TGS$ and $TGS_{R}$ in gases (a) $Ar:CO_2$ = 90:10 and (b) $Ar:iC_4H_{10}$ = 95:5}
	\label{spec-sinp}
\end{figure}

\begin{figure}[htbp]
    \centering
    \includegraphics[width=9.0cm]{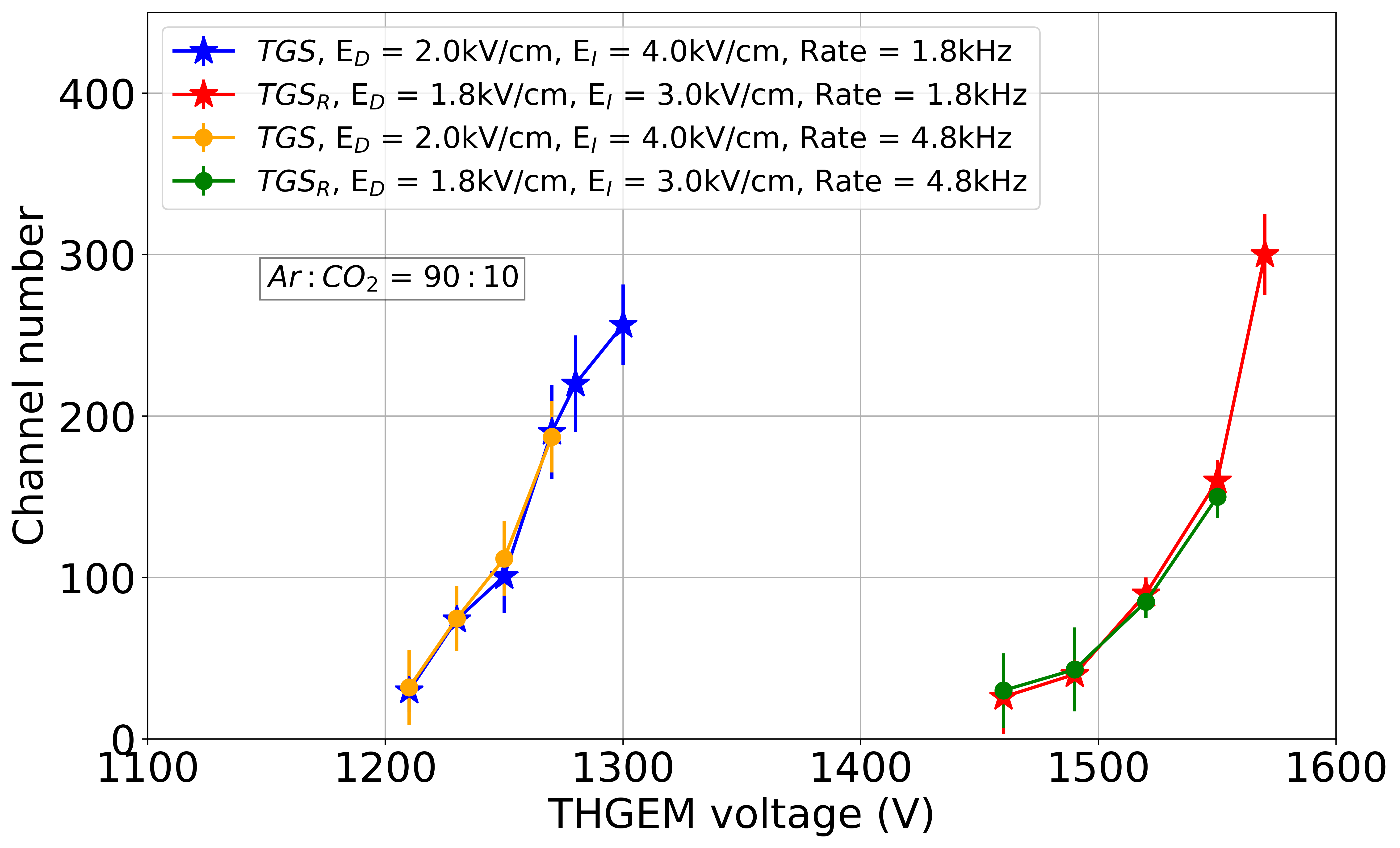}
    \caption{Operating range in argon-carbon dioxide mixture}
    \label{Oprange_9010}
\end{figure}

Figure~\ref{spec-sinp} shows the $^{55}$Fe spectra for both $TGS$ and $TGS_{R}$ in the two argon based gas mixtures. The next step has been to measure the optimum working voltage range for both the THGEMs. Figure~\ref{Oprange_9010} shows the operating range for $TGS$ and $TGS_{R}$ in argon-carbon dioxide gas mixture. It is observed that working voltage range is different for both the local THGEMs . $TGS_{R}$ has a working voltage range on the higher side. The reason is the presence of rims in $TGS_{R}$ which decreases the ampliplication field inside the holes and hence, shifts the working voltage range to the higher values. Another important point is that the voltage range for $TGS_{R}$ is slightly larger than that for $TGS$. This is again due to the presence of rims in $TGS_{R}$ which reduces the discharges at higher voltages.  

\begin{figure}[htbp]
	\centering
	\includegraphics[width=9.0cm]{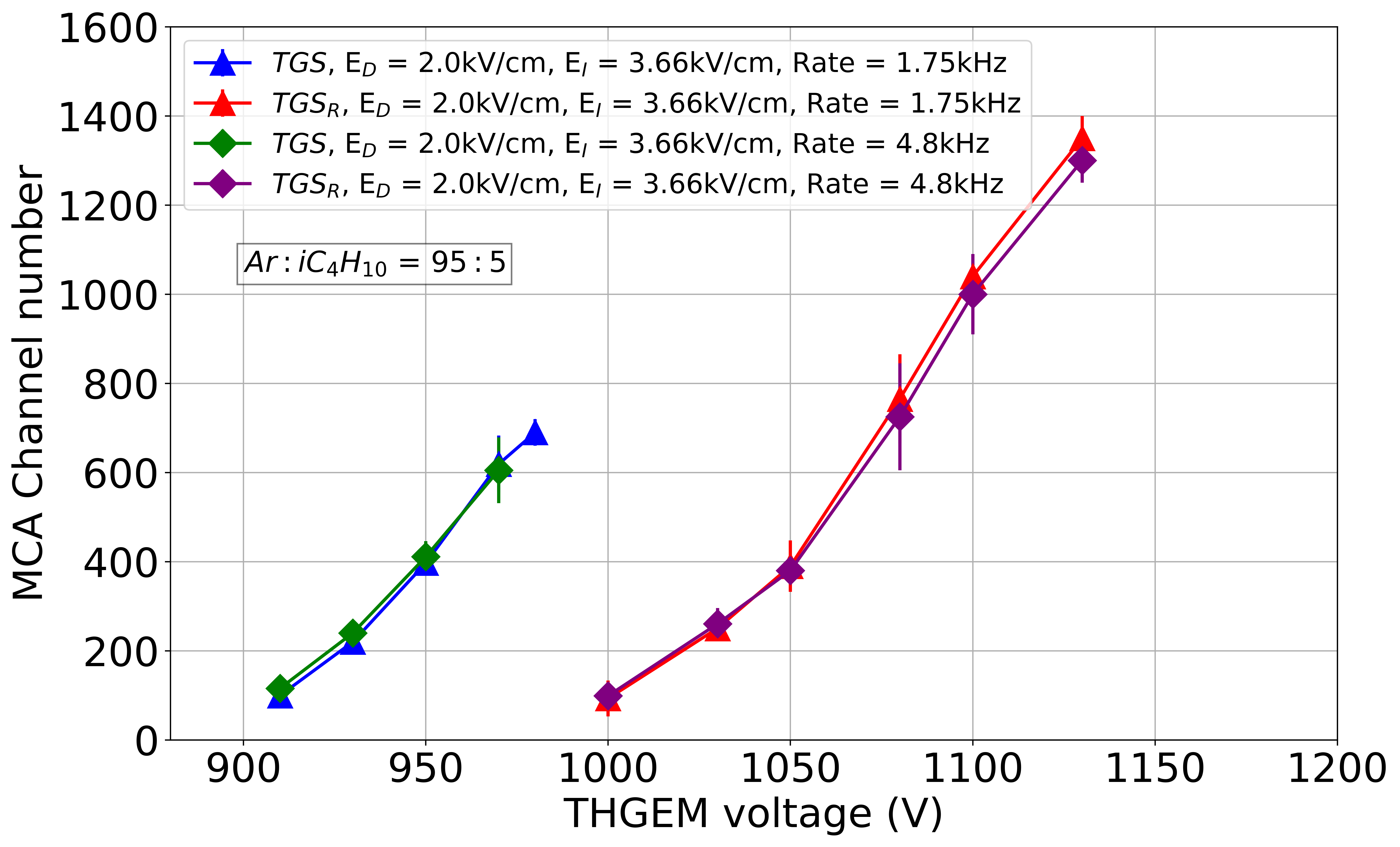}
	\caption{Operating range in argon-isobutane mixture}
	\label{Oprange_955}
\end{figure}
Furthermore, using two different rates of source also makes a difference in the allowed voltage range. Voltage range is less for both $TGS$ and $TGS_{R}$ when a high-rate source (4.8kHz) is used in argon-carbon dioxide mixture. On testing the operating voltages for $TGS$ and $TGS_{R}$ in argon-isobutane mixture, it is observed that $TGS_{R}$ is able to withstand gain higher than $TGS$ as shown in figure~\ref{Oprange_955}. In addition, on increasing radiation rate of the source, there is no discharge observed for $TGS_{R}$ and it is found to have the same operating range for both low-rate and high-rate source. On the contrary, $TGS$ could not withstand the same voltage range with high-rate source as it did with low-rate source. In other words, the working voltage range for $TGS$ is observed to decrease on increasing the radiation rate of the source in argon-isobutane mixture as well. This is due to the absence of the rim in $TGS$ which makes it more prone to discharges and thus, decreases the working voltage range.

\begin{figure}[htbp]
	\centering
	\begin{subfigure}{0.49\textwidth}
		\centering
		\includegraphics[width=\textwidth]{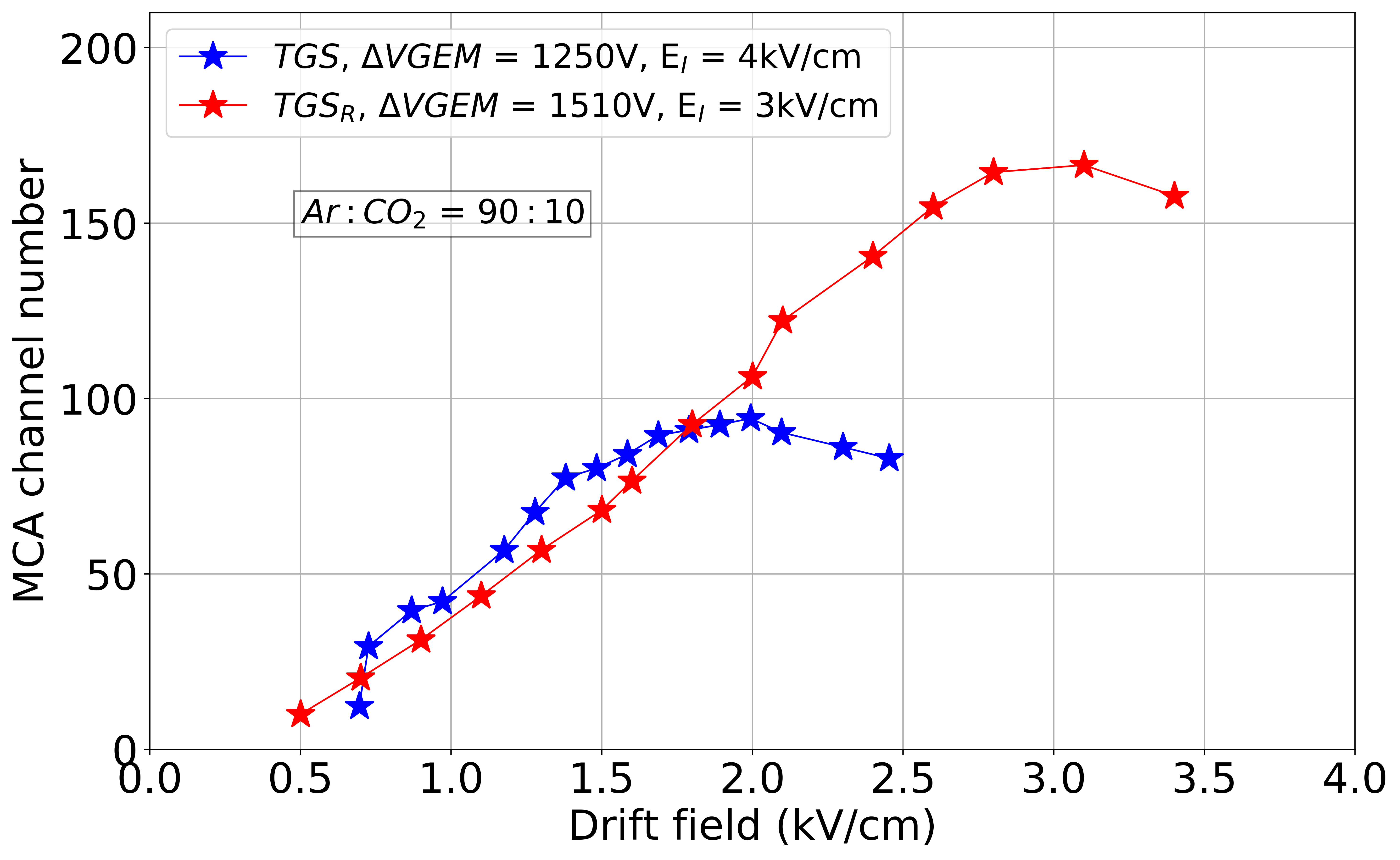}
		\caption{}
		\label{coll-9010}
	\end{subfigure}
	\hfill
	\begin{subfigure}{0.49\textwidth}
		\centering
		\includegraphics[width=\textwidth]{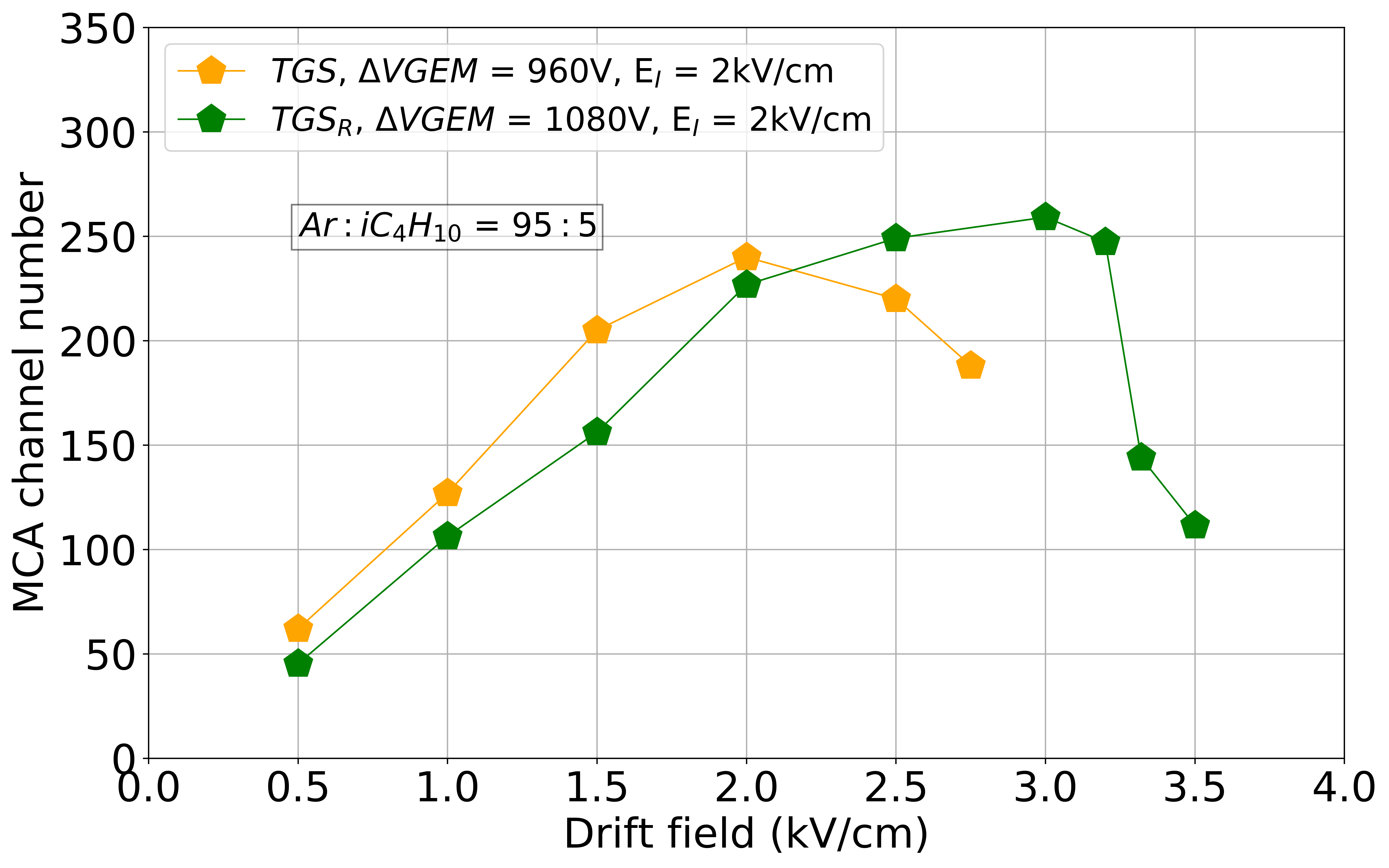}
		\caption{}
		\label{coll-955}
	\end{subfigure}
	\caption{Variation of gain with drift field for (a) argon-carbon dioxide and (b) argon-isobutane gas mixtures}
	\label{coll-sinp}
\end{figure}

To optimize the drift field and induction field values for gain measurement, mean of the fitted MCA photo peaks have been recorded by varying the drift field and induction field values while keeping the $\Delta$VGEM constant. In both the cases, drift and induction field values have not been increased beyond a value which can initiate an avalanche in the concerned gas. 

 Figure~\ref{coll-9010}  shows the variation of MCA channel number with $E_D$ at constant values of $\Delta$VGEM and $E_I$ for argon-carbon dioxide based gas mixture. It has been observed that channel number increases with $E_D$, reaches a maximum (plateau-like region) and sligtly decreases on further increasing $E_D$. The value of $E_D$ corresponding to the maximum gain is different for both $TGS$ and $TGS_{R}$. The gain is maximum at $E_D$ close to 2kV/cm and 3.125kV/cm for $TGS$ and $TGS_{R}$ respectively. This difference is due to the different ratio of field values at the entrance of holes of both the THGEMs.
 
Similarly, for argon-isobutane based gas mixture(figure~\ref{coll-955}), variation of MCA photopeak channel number has been recorded for different $E_D$ keeping the other two parameters $\Delta$VGEM and $E_I$ fixed. Gain is found to increase, reach a maximum and then decrease finally.  Once again the value of $E_D$ corresponding to the maximum gain has been found to be different for $TGS$ and $TGS_{R}$.

\begin{figure}[htbp]
	\centering
	\begin{subfigure}{0.49\textwidth}
		\centering
		\includegraphics[width=\textwidth]{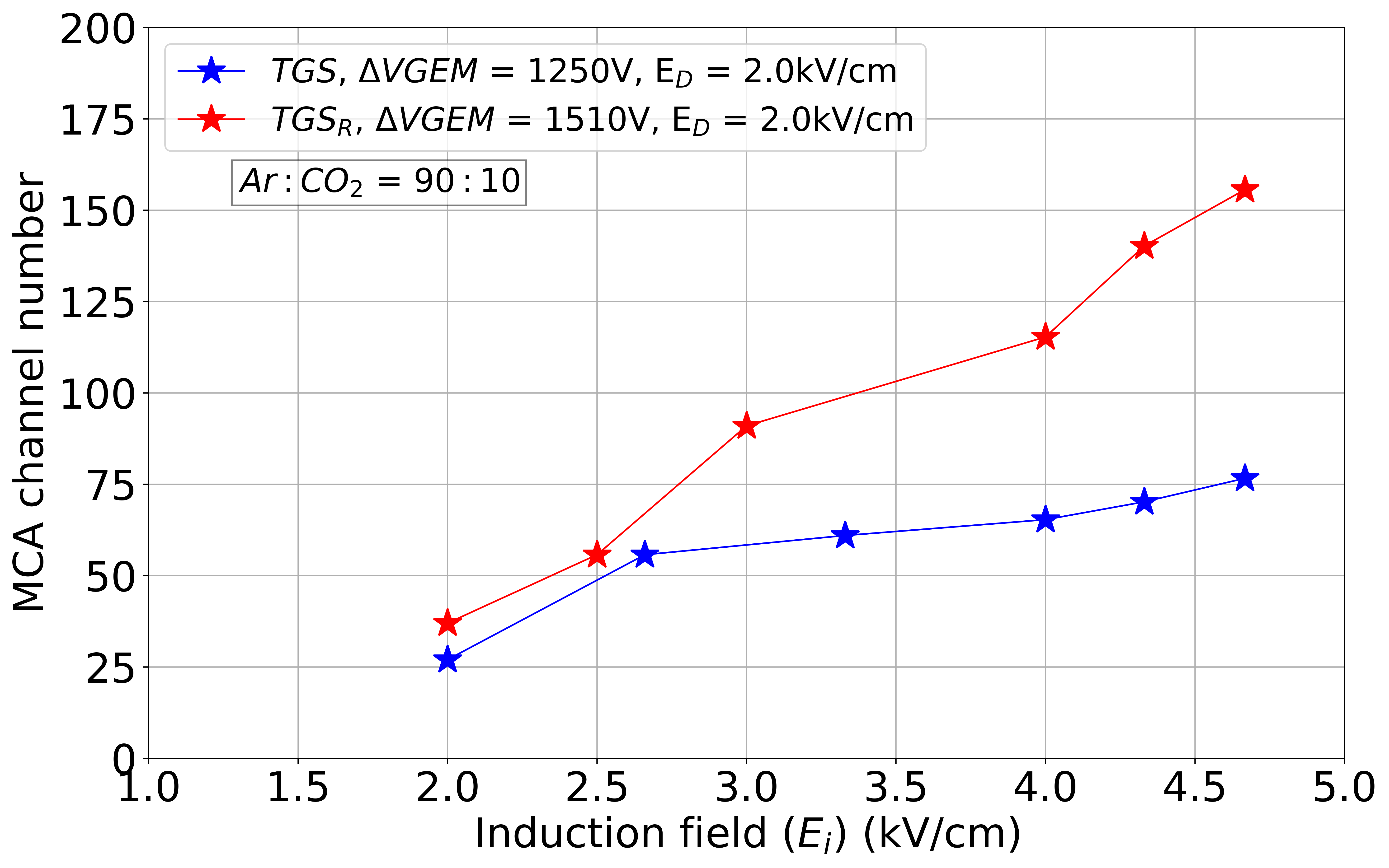}
		\caption{}
		\label{ind-9010}
	\end{subfigure}
	\hfill
	\begin{subfigure}{0.49\textwidth}
		\centering
		\includegraphics[width=\textwidth]{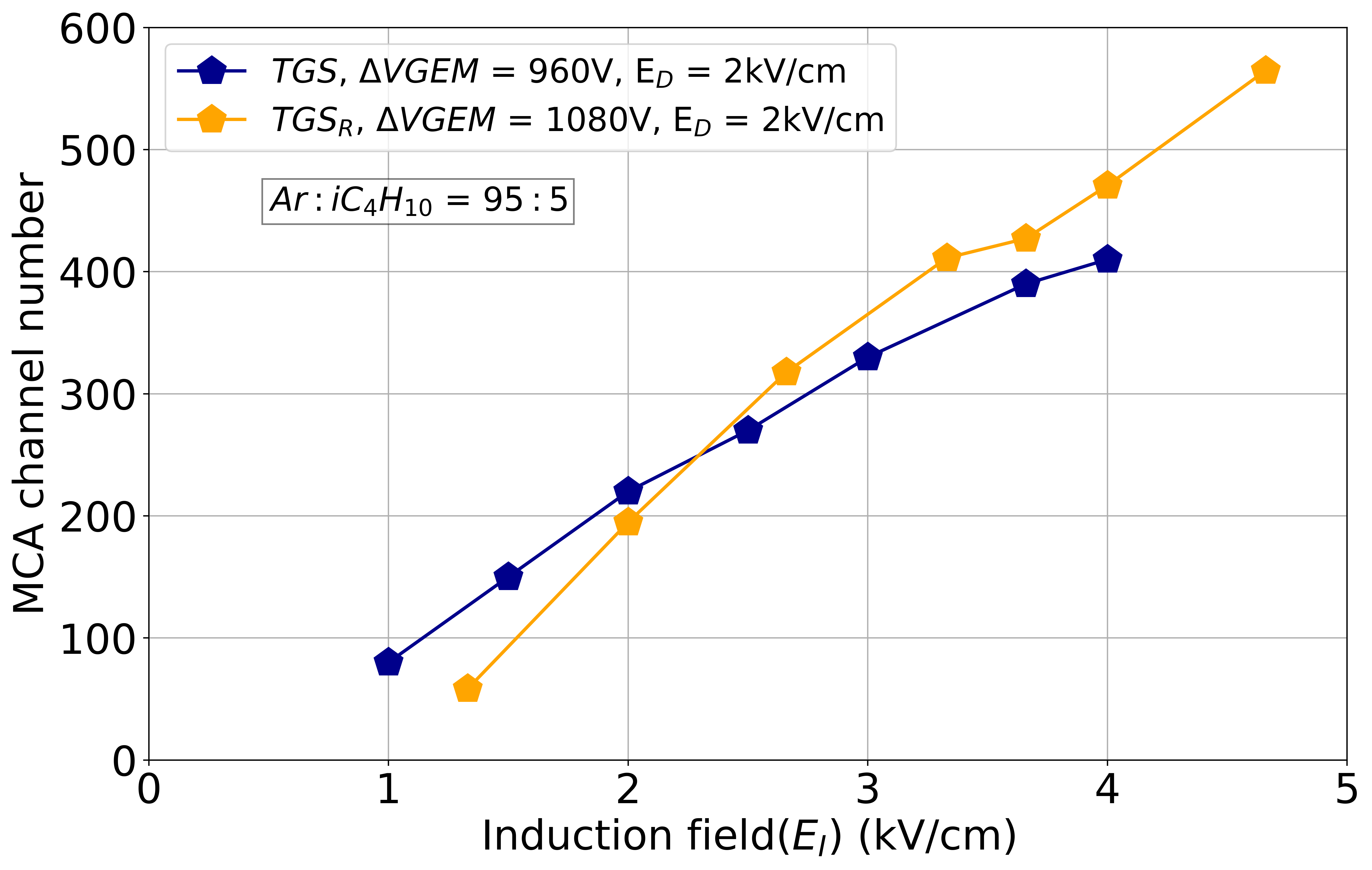}
		\caption{}
		\label{ind-955}
	\end{subfigure}
	\caption{Variation of gain with induction field in (a) argon-carbon dioxide and (b) argon-isobutane mixtures}
	\label{IndV_sinp}
\end{figure}

Variation of gain with induction field ($E_I$) have been studied keeping the $\Delta$VGEM and $E_D$ fixed. Figure~\ref{IndV_sinp} shows that the gain increases with increasing induction field for both $TGS$ and $TGS_{R}$ in both the argon based gas mixtures.

\subsection{Gain measurement}

For measuring the gain of the detectors, drift and induction fields have been kept fixed at the values around which the effective gain is close to maximum. The desired voltage at which the gain is to be measured, has been applied to the detector and kept undisturbed till the time the THGEM PCBs are fully polarized. This dielectric polarization time is found to be different for different THGEMs depending on their geometry. Once the polarization is complete at a given THGEM voltage, the detector is irradiated with a source and large number of MCA spectra have been saved till the time the photo peak position in the MCA saturated. These 5.9 keV x-ray spectra have been fitted with a Gaussian function. The area under such a gaussian distribution has been calculated which is equal to the total number of counts for that measurement. Finally, the equivalent rate of irradiation for each measurement has been estimated by dividing the area of the spectra by the duration of each measurement. Upon the saturation of photo peak position, current from readout is measured using a pico-ammeter. Current data for 5 minutes have been saved and fitted with a gaussian to obtain the average value of saturation current.
Gain is calculated using the formula 

\begin{equation}
\label{eq:gain}
\begin{split}
G = I/(n_{p}\times R\times e) \,,
\end{split}
\end{equation}
where 'G' is the effective gain, 'n$_{p}$' is the mean number of primaries obtained from HEED simulation, R is the rate of the source used and 'e' is the electronic charge.

\begin{figure}[htbp]
    \centering
    \includegraphics[width=9.05cm]{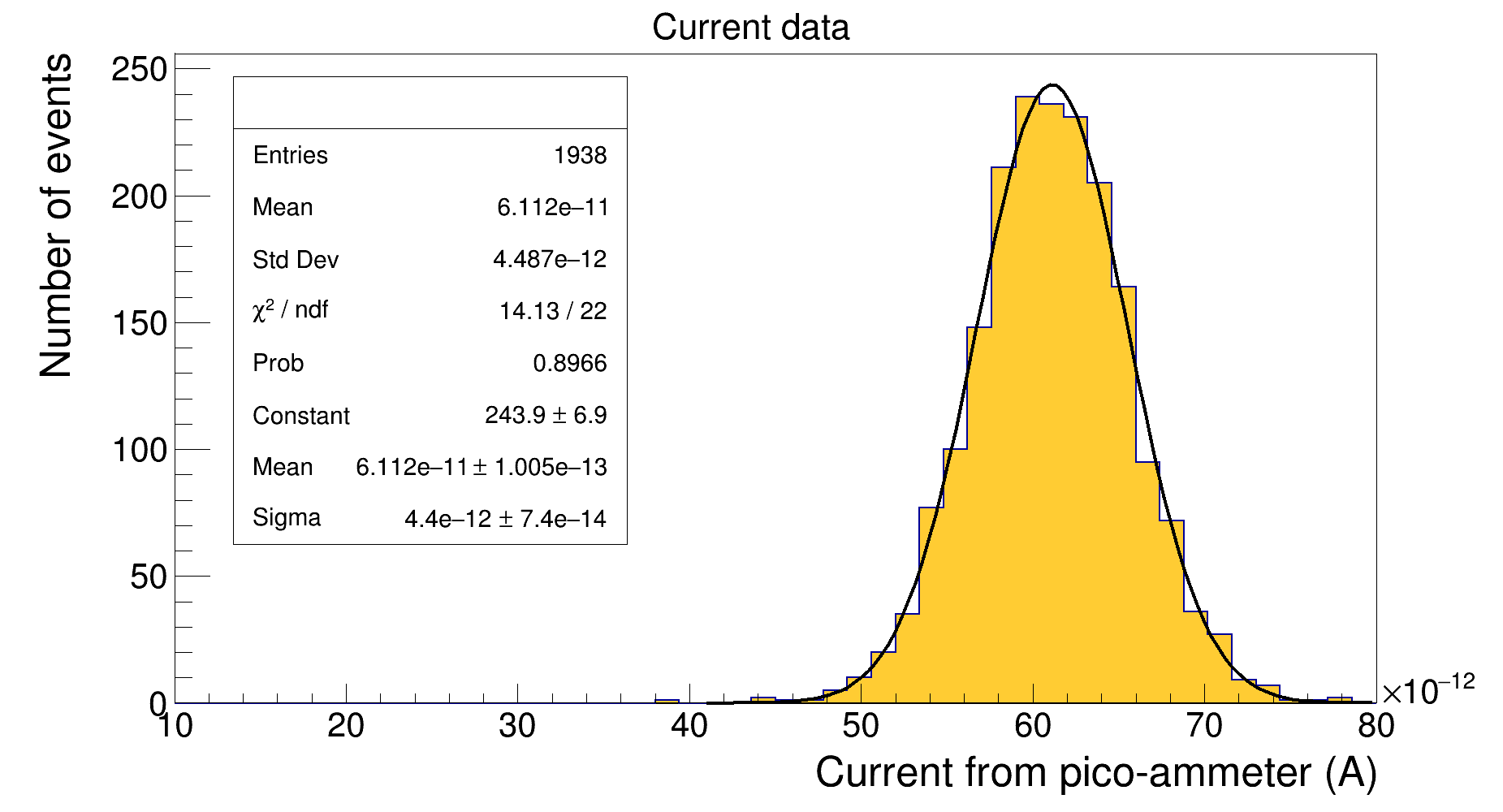}
    \caption{Current measurement using pico-ammeter}
  \label{picoamm-data}
\end{figure}

\begin{figure}[htbp]
	\centering
	\includegraphics[width=8.0cm]{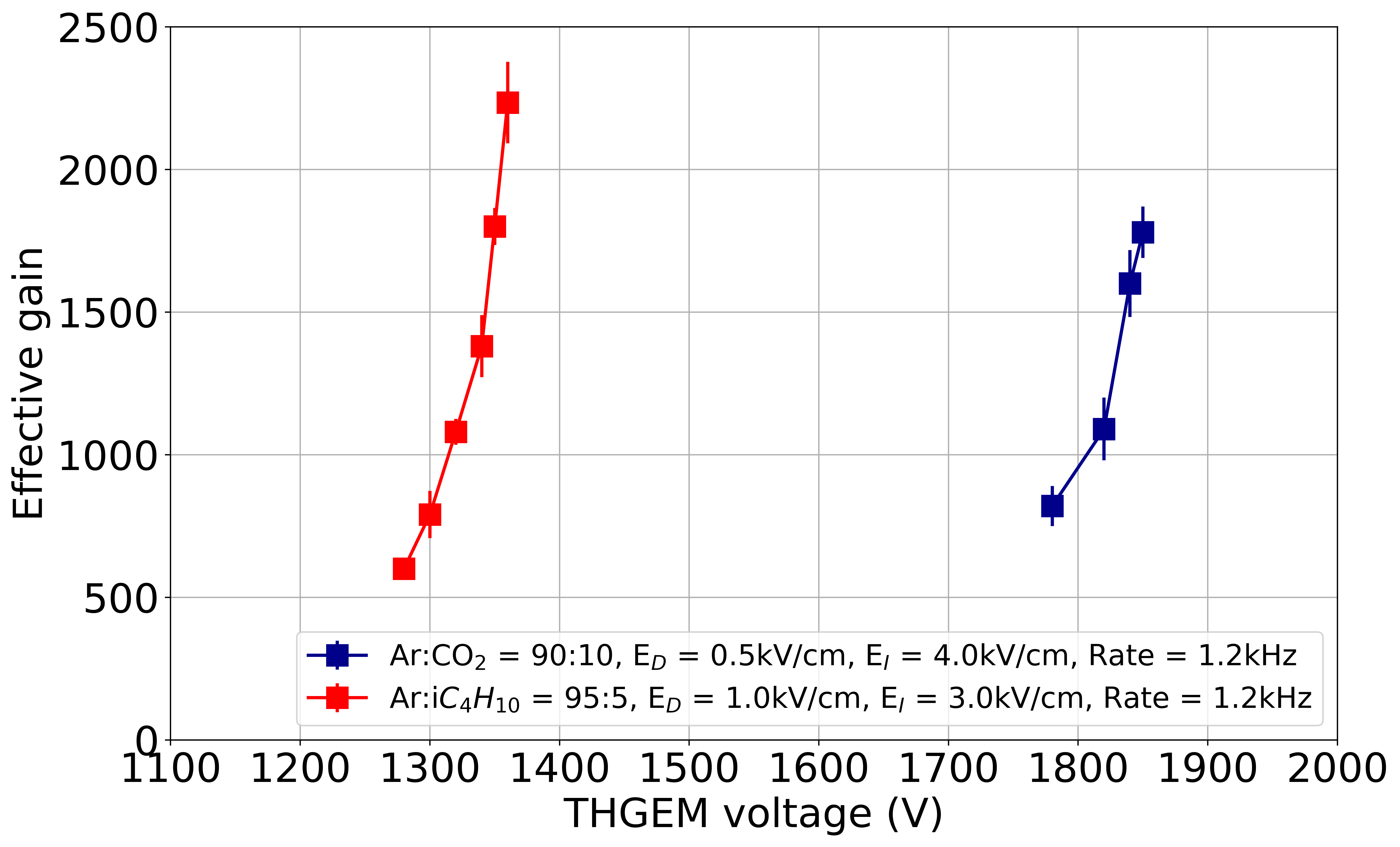}
	\caption{Effective gain values for $TGC_{R}$ in argon-carbon dioxide and argon-isobutane mixture with a source of rate 1.2kHz}
	\label{gain_cern}
\end{figure}

 A typical readout current data measured using pico-ammeter is shown in the figure~\ref{picoamm-data}. Mean value of the gaussian fitted photopeak is used for the gain measurement. Effective saturated gain once the radiation charging up process gets over for $TGC_{R}$, $TGS$ and $TGS_{R}$ have been measured using the formula given in equation~\ref{eq:gain}. Figure~\ref{gain_cern} shows the gain curve for $TGC_{R}$ in argon based mixtures.
 
 \begin{figure}[htbp]
 	\centering
 	\includegraphics[width=8.0cm]{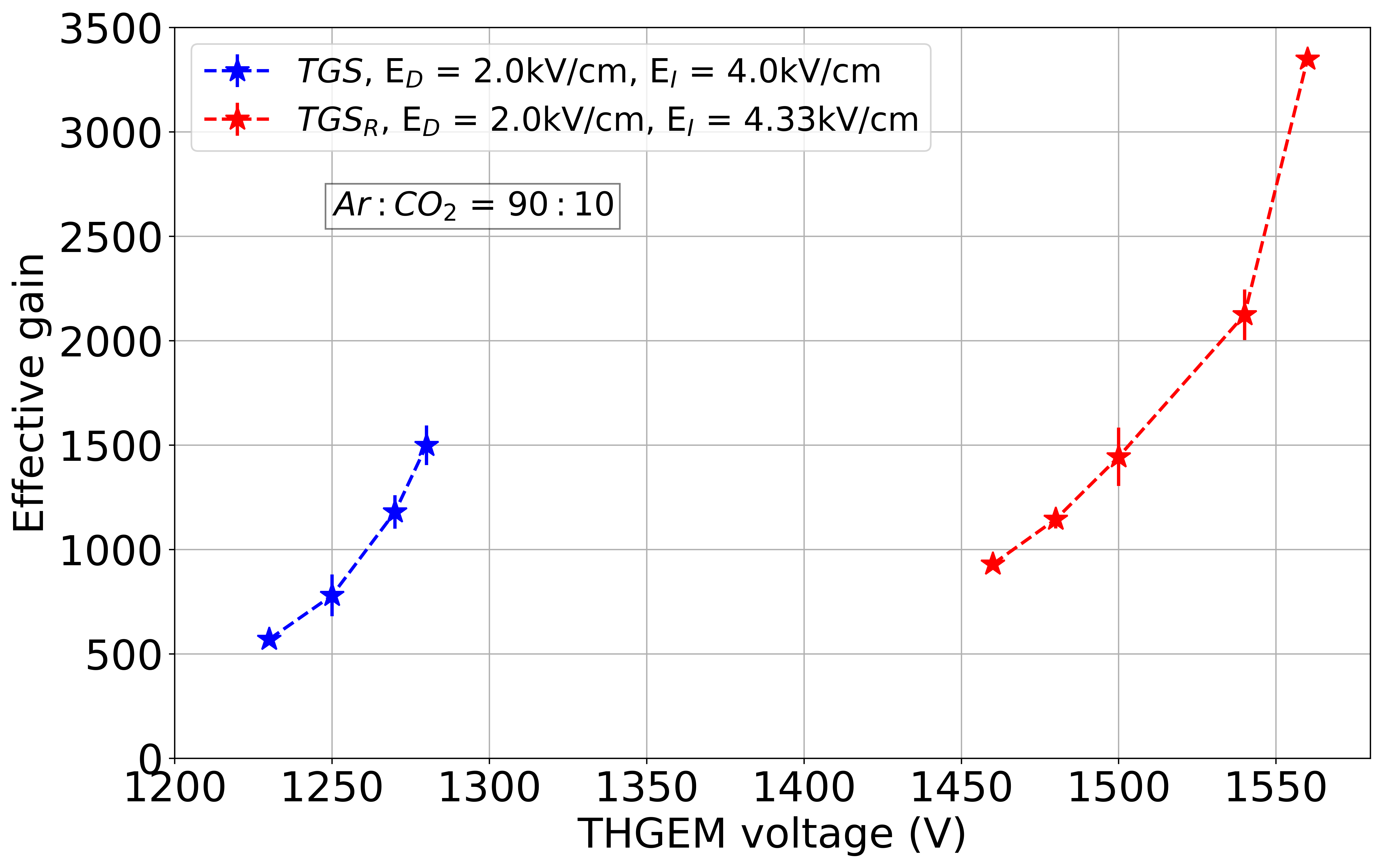}
 	\caption{Variation of gain for $TGS$ and $TGS_{R}$ in argon-carbon dioxide mixture}
 	\label{gain-ar90}
 \end{figure}
 
 \begin{figure}[htbp]
 	\centering
 	\includegraphics[width=8.0cm]{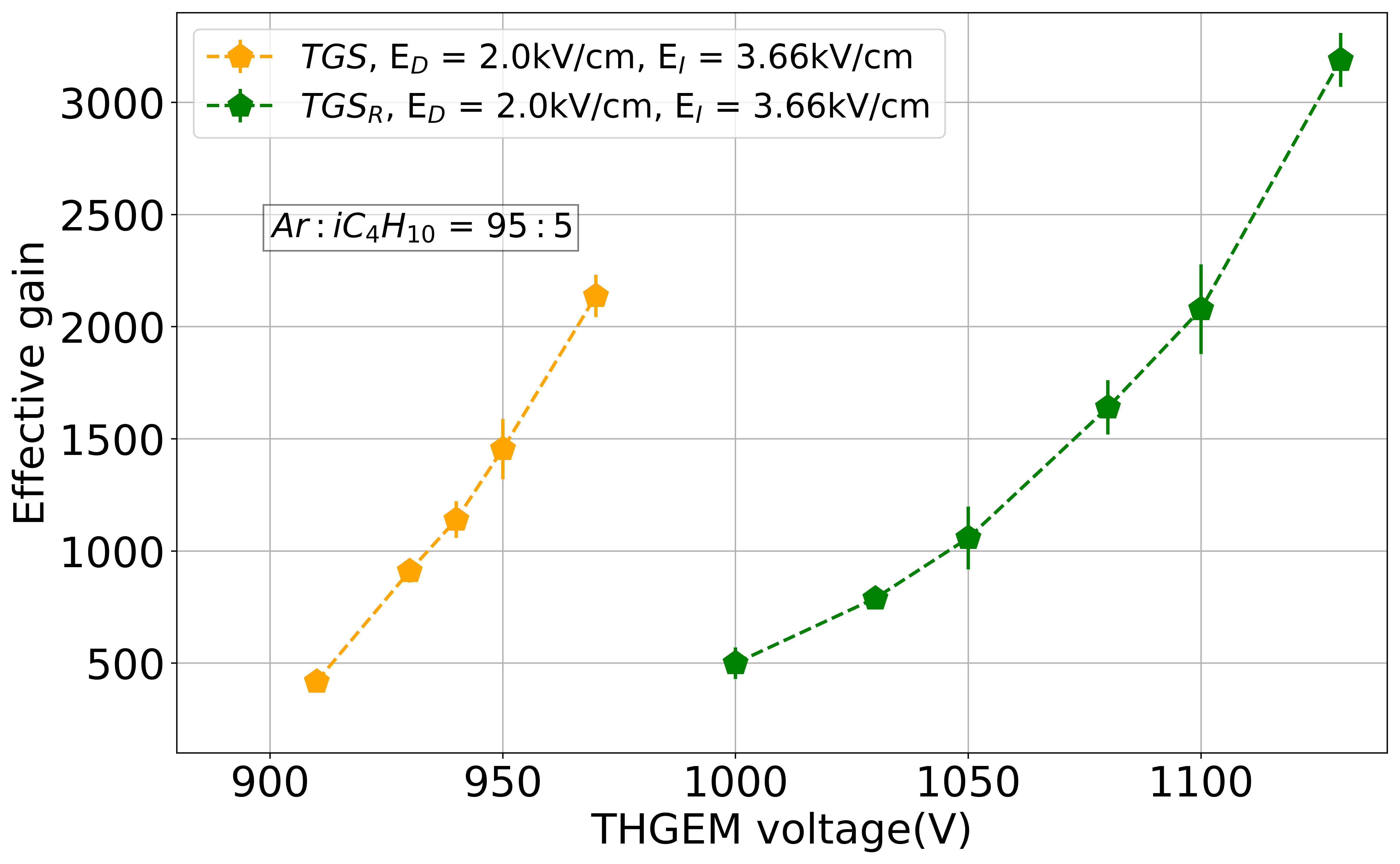}
 	\caption{Variation of gain for $TGS$ and $TGS_{R}$ in argon-isobutane mixture}
 	\label{gain-ariso}
 \end{figure}

Similarly, effective gain values for locally fabricated THGEMs in argon-carbon dioxide and argon-isobutane gas mixtures are shown in figure~\ref{gain-ar90} and figure~\ref{gain-ariso} respectively. It is observed that the working voltage range is higher in $TGS_{R}$. In addition to a larger working voltage range, $TGS_{R}$ also yields higher effective gain value. The reason could be higher transmission efficiency for $TGS_{R}$ as compared to $TGS$.
Similar results have been observed while measuring effective gain for local THGEMs in argon-isobutane gas mixture as plotted in figure~\ref{gain-ariso}. Gain for $TGS_{R}$ has been found to be higher as compared to $TGS$.

	

\subsection{Energy resolution}
$^{55}Fe$ spectrum obtained during gain measurement for each applied THGEM voltage is fitted with a gaussian function. The mean and sigma values of the fitted gaussian photopeak have been used to calculate the energy resolution at that voltage. Energy resolution is defined to be
\begin{equation}
\label{eq:2}
\begin{split}
E_{res} = FWHM/E = 2.355 \times \sigma_E/E
\end{split}
\end{equation}
where 'E' and '$\sigma_E$' are the mean and sigma of the gaussian fitted MCA spectrum.

\begin{figure}[htbp] 
	\centering
	\includegraphics[width=8.0cm]{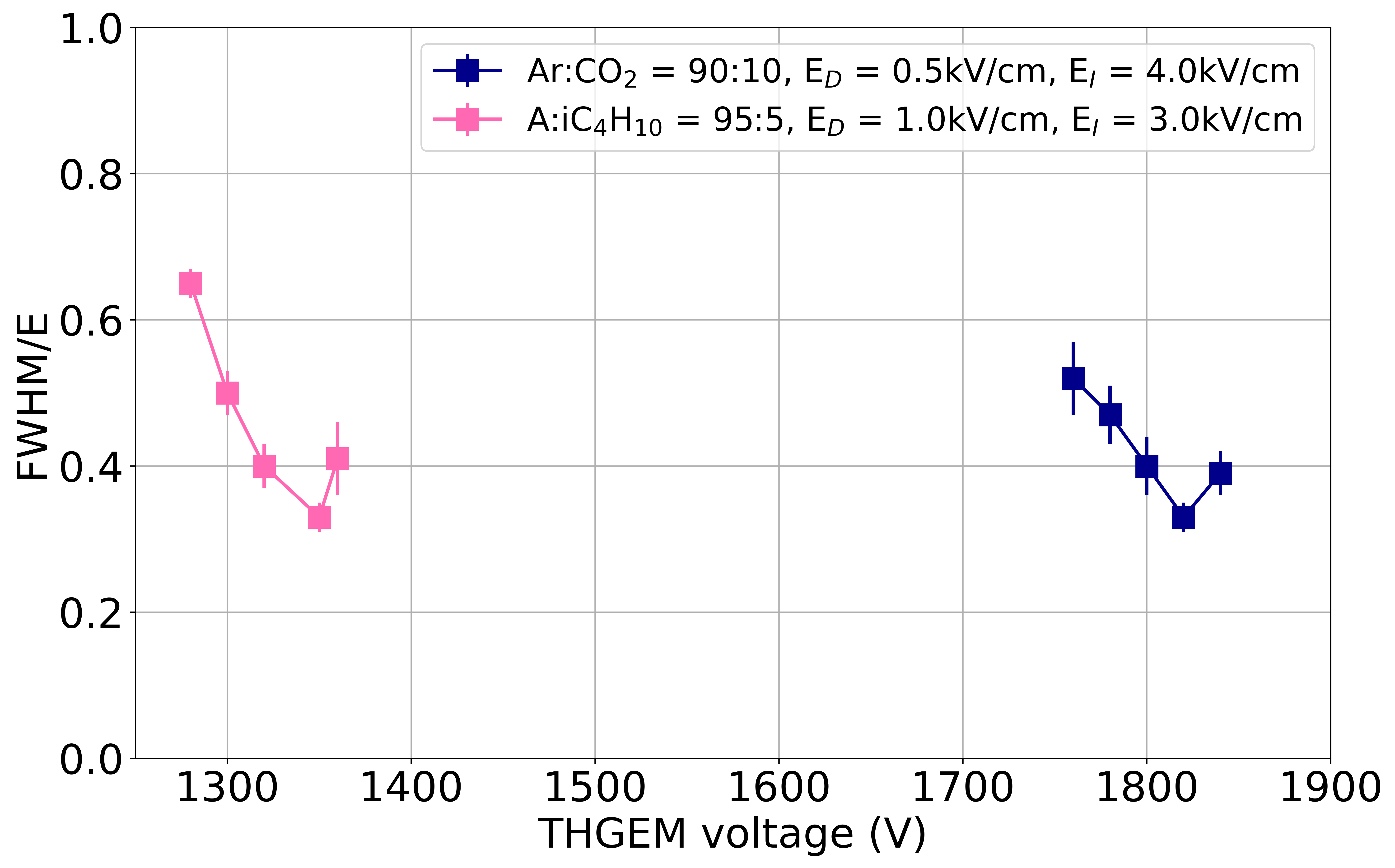}
	\caption{Variation of energy resolution with the applied THGEM voltage for $TGC_{R}$}
	\label{energyres_cern}
\end{figure}

These energy resolution values have been calculated once the gain is found to saturate for a partcular applied THGEM voltage. $E_{res}$ value is found to decrease on increasing the THGEM voltage, reach a minimum value and then again it increases on further increasing the voltage across THGEM. The same variation has been observed for all the three THGEMs. Figure~\ref{energyres_cern} shows the variation of energy resolution with applied voltage for $TGC_{R}$ in both argon-carbon dioxide and argon-isobutane gas mixtures. The best resolution obtained fot $TGC_{R}$ in both the gases is around 32-33\%. This value of energy resolution is found to be affected by the applied drift and induction fields. 

\begin{figure}[htbp]
	\centering
	\begin{subfigure}{0.49\textwidth}
		\centering
		\includegraphics[width=\textwidth]{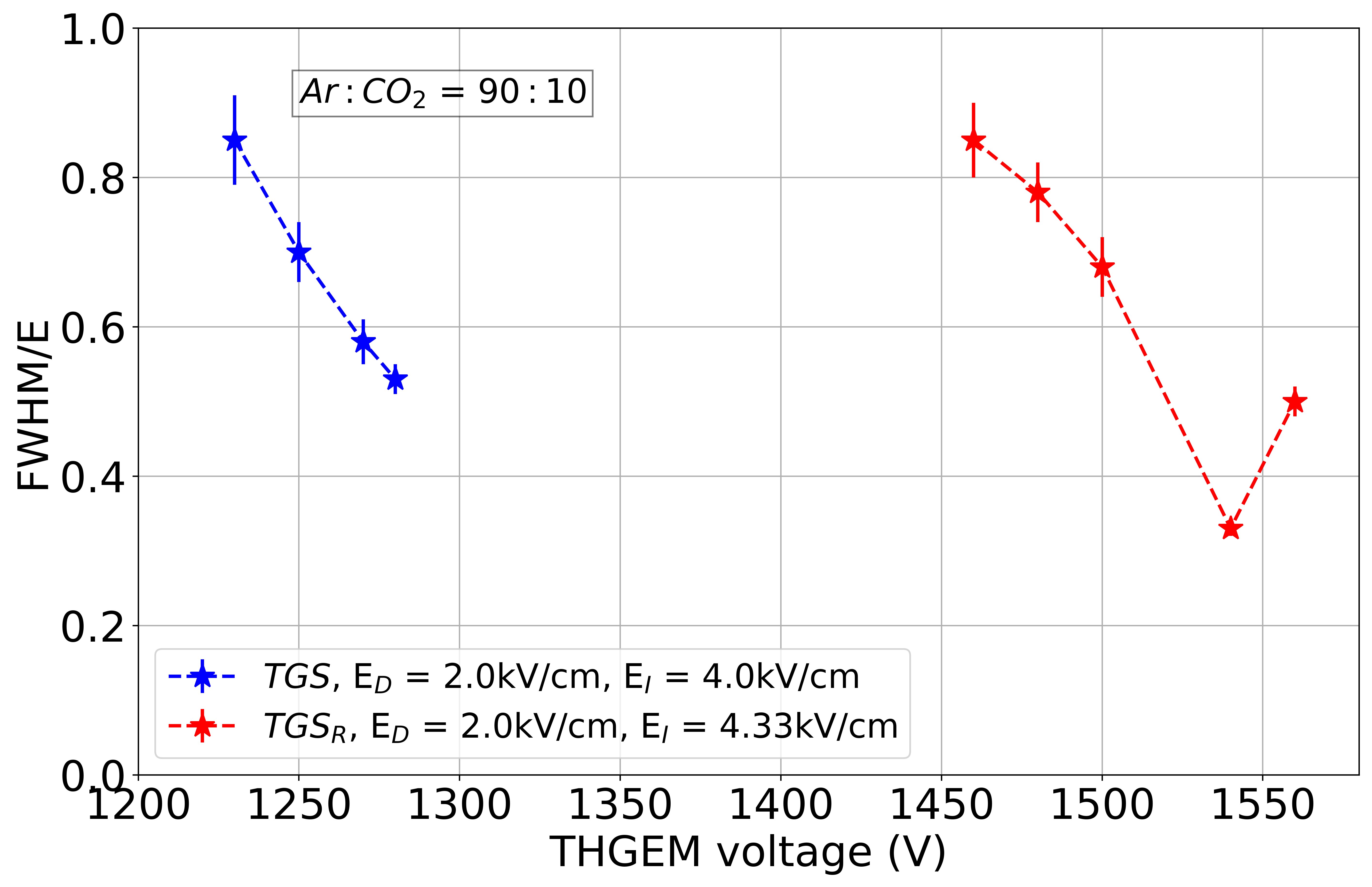}
		\caption{$Ar:CO_2$ = 90:10}
		\label{ener-9010}
	\end{subfigure}
	\hfill
	\begin{subfigure}{0.49\textwidth}
		\centering
		\includegraphics[width=\textwidth]{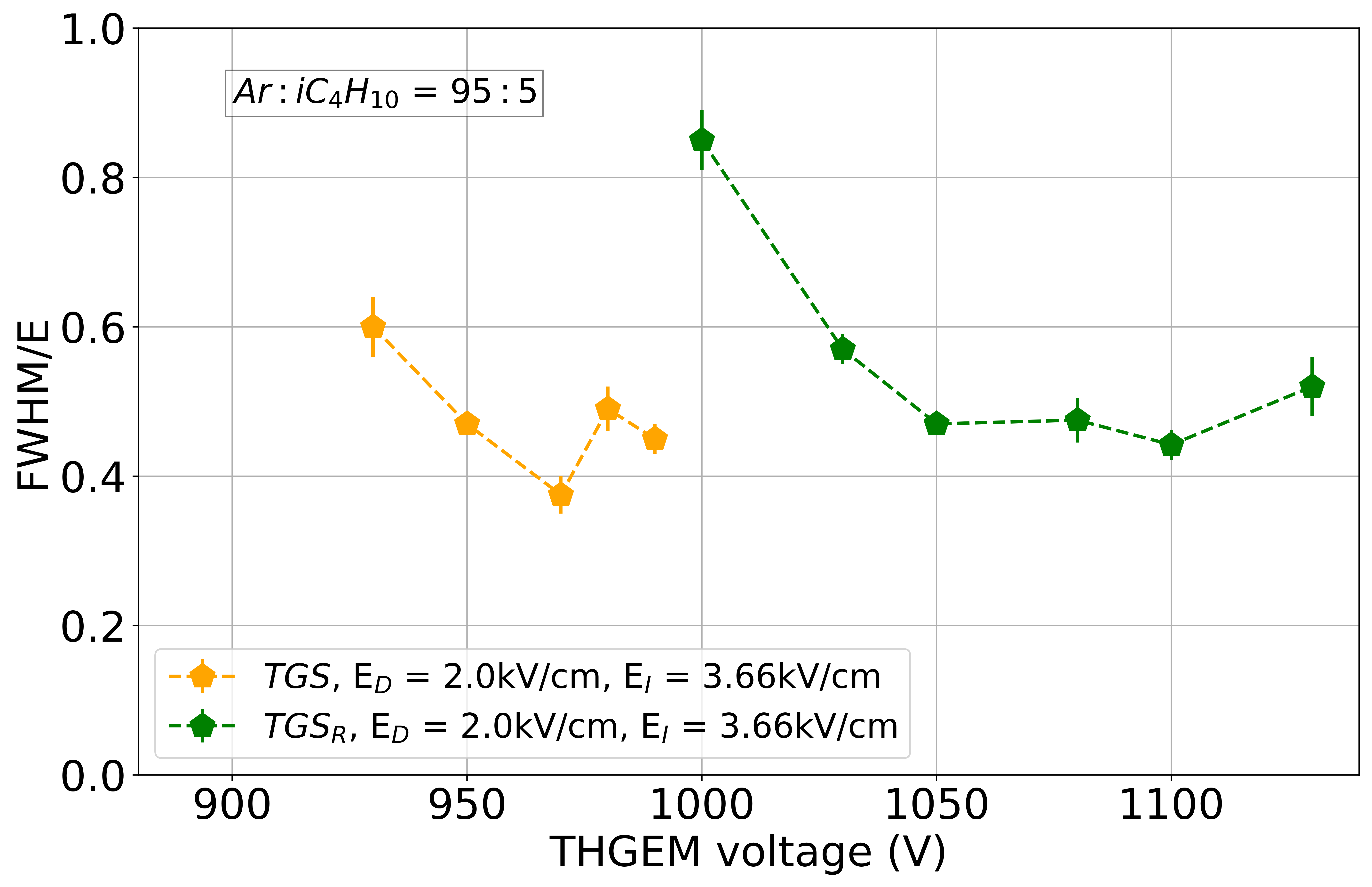}
		\caption{$Ar:iC_4H_{10}$ = 95:5}
		\label{ener-955}
	\end{subfigure}
	\caption{Variation of energy resolution for local THGEMs with the applied voltage in (a) argon-carbon dioxide and (b) argon-isobutane mixtures}
	\label{ener_sinp}
\end{figure}

Similarly, the energy resolution values for $TGS$ and $TGS_{R}$ in both the argon based gas mixtures are shown in figure~\ref{ener_sinp}.There are two points to be discussed in the context of energy resolution. First is the presence and absence of rim affecting the energy resolution, second is the gas mixture. In argon-carbon dioxide gas mixture, $TGS_{R}$ is observed to have a better resolution for higher voltage points than $TGS$. $TGS$ fails to yield better resolution than 52\% due to it having discharges at higher voltage points. However in argon-isobutane mixture, $TGS$ seems to have slightly better energy resolution than $TGS_{R}$. This is due to the quenching property of isobutane which reduces discharges at higher voltages. Furthermore, $TGS_{R}$ is observed to yield a decent resolution fluctuating around 45\% for a larger voltage range. This can be attributed to the presence of rim in $TGS_{R}$ and a quencher gas (isobutane) in the studied gas mixture. This combination of rim and isobutane increases the voltage range for $TGS_{R}$ and a saturation valley is observed around 45\%. Finally, the overall energy resolution for both the THGEMs is expected to improve by tuning the drift and induction field values.

\section*{Conclusion}
An elaborate simulation study for THGEMs of different geometries has been performed to optimize the size of the rim in order to maximize the electron transmission. Collection and transmission efficiencies are found to increase with increasing size of the rim and are found to be maximum for the rim size of 80-100$\mu m$. To take into account the hole-rim misalignment, rim-offset study has been carried out which shows that an offset within 10-15$\mu m$ does not affect the electron transparency of THGEMs. Finally, simulating two extreme cases, one with no rim and another with 120$\mu m$, shows that electron transmission efficiency is almost double for the one with rim of 120$\mu m$. Drawing conclusions from the simulation results and to see whether the same is reflected in the experiments, THGEMs with and without rims have been fabricated locally in SINP, tested and characterized using $^{55}$Fe source in argon based gas mixtures. Their performance have been compared with the performance of THGEM from CERN ($TGC_{R}$). 

Operating voltage range for $TGS$ and $TGS_{R}$ are found to have no overlap regions. Voltage range for $TGS_{R}$ is found to shift towards the higher magnitudes, which can be attributed to the presence of rims. Electric field inside the holes decreases due to the presence of rim, as a result of which voltage range for $TGS_{R}$ shifts towards higher magnitudes than $TGS$ in both the gases. Moreover, operating voltage range is found to depend on the rate of irradiation of source. Presence of rim is found to reduce discharges at higher voltages for high rate sources, thus allowing higher working voltage range. Furthermore, these THGEMs, specially $TGS$, is found to perform better in argon-isobutane gas mixture.

Gain of more than $3.2 \times 10^3$ have been achieved with $TGS_{R}$. The presence of the etched rim around the THGEM holes, is found to be essential for reducing discharge-occurrence probability significantly. This permitted operation at higher voltages and at higher detector gains. $TGS$ is found to have a maximum gain of $2.2 \times 10^3$ in argon-isobutane gas mixture which is better than the maximum gain obtained in argon-carbon dioxide gas mixture. In terms of energy resolution, $TGS$ is found to yield slightly better resolution $TGS_{R}$ in argon-isobutane mixture, but the latter yields a satisfactorily good resolution for a larger voltage range.

To conclude, $TGS_{R}$ performs better close to the discharge limit by significatly reducing discharges and allowing operation at higher gains. Furthermore, it has excellent electron transmission close to 90\%. Presence of a quencher gas improves the operation of $TGS$. Finally both $TGS$ and $TGS_{R}$ yield reasonably good energy resolution. In future, we plan to use THGEMs in double stage configuration and study their position resolution for muon tracking purposes.

\section*{Acknowledgement}
Authors would like to acknowledge their respective Institutes and Universities for the funding, laboratory and computational facilities to carry out the investigations and simulation. They would like to thank Mr. Biswajit Majumdar and Mr. Samarjit Majumdar for helping them in the fabrication process. Authors would also like to thank Mr. Subhendu Das, Mrs. Lipi Das and Mr. Arindam Das for their help in the THGEM gerber file design. Finally they would like to thank Prof. Chinmay Basu for his support. This work has partially been performed in the framework of the RD51 Collaboration and they would like to acknowledge the members of the RD51 Collaboration for their help and suggestions.

%




\end{document}